\documentclass[12pt]{article}
\usepackage{amsfonts}
\usepackage{eurosym}
\usepackage{graphicx, tikz, tikzscale}
\usepackage{booktabs}
\usepackage{float}
\usepackage{caption, subcaption}
\usepackage{amssymb}
\usepackage{comment}
\usepackage{amsmath, bbm, amsthm,adjustbox}
\usepackage{booktabs}
\usepackage{multirow}
\usepackage{enumerate,multicol,enumitem}
\usepackage{color}
\usepackage{tikz-network}
\usepackage[authoryear]{natbib}
\usepackage{hyperref}
\usepackage{setspace}
\usepackage[title]{appendix}
\usetikzlibrary{positioning}

\hypersetup{colorlinks=true, linkcolor=red, citecolor=blue}

\newtheorem{lemma}{{\bf \sc Lemma}}

\newtheorem{corollary}{{\bf \sc Corollary}}
\newtheorem{proposition}{{\bf \sc Proposition}}

\def\eproof{\hbox{\hskip3pt\vrule width4pt height8pt depth1.5pt}}

\begin{document}
\title{
Credit Freezes, Equilibrium Multiplicity, and Optimal Bailouts in Financial Networks.
}
\author{Matthew O. Jackson and Agathe Pernoud\thanks{%
Department of
Economics, Stanford University, Stanford, California 94305-6072 USA.
Jackson is also an
external faculty member of the Santa Fe Institute.
Email:  jacksonm@stanford.edu and agathep@stanford.edu.  We gratefully
acknowledge financial support under NSF grants SES-1629446 and SES-2018554.
We thank Marco Bardoscia, Philip Bond, Celso Brunetti, Ozan Candogan, Matt Elliott, Gerardo Ferarra,
David Hirshleifer, Arjun Mahalingam, Carlos Ramirez, Tarik Roukny, Alireza Tahbaz-Salehi, Erol Selman, and especially
Co-Pierre Georg, for
helpful conversations and comments.  We also thank the editor Itay Goldstein, and the anonymous referees for suggestions that substantially improved the paper.
This paper incorporates some results from
``Distorted Investment Incentives, Regulation, and Equilibrium Multiplicity in a Model of Financial Networks.'' We split that paper into this one and another,
Jackson and Pernoud \citeyearpar{jacksonp2021b} (SSRN: \url{https://papers.ssrn.com/abstract=3311839}). Both papers contain new results beyond what appeared in the original paper.
}}
\date{Draft: May 2023 
}
\maketitle

\begin{abstract}
We analyze how interdependencies in financial networks can lead to self-fulfilling insolvencies and multiple possible equilibrium outcomes.  We show that multiplicity arises if and only if there exists a certain type of dependency cycle in the network, and characterize banks' defaults in any equilibrium.
We use this analysis to show that finding the cheapest bailout policy that prevents self-fulfilling insolvencies is computationally hard (and hard to approximate), but that the problem has intuitive solutions in specific network structures.  Bailouts have an indirect value as making a bank solvent improves its creditors' balance sheets and reduces their bailout costs. We show that an algorithm that leverages these indirect benefits ensures systemic solvency at a cost that never exceeds half of the overall shortfall.  In core-periphery networks, it is often optimal to bail out peripheral banks first as opposed to targeting core banks directly.


\textsc{JEL Classification Codes:}  D85, F15, F34, F36,
F65, G15, G32, G33, G38

\textsc{Keywords:} Financial Networks, Systemic Risk, Financial Crisis, Credit Freeze, Default Risk, Financial Interdependencies, Bailouts, Equilibrium Multiplicity
\end{abstract}

\setcounter{page}{0}\thispagestyle{empty} \newpage

\section{Introduction}

Today's financial sector is characterized by strong interdependencies,
with large amounts of capital circulating among financial firms.
For instance, Duarte and Jones \citeyearpar{duartej2017} estimate that 23\% of the assets of bank holding
companies come from other financial intermediaries, as well as 48\% of their
liabilities - {\sl almost half}.\footnote{See their Tables 1 and 2.  The difference reflects the fact that many
other types of financial institutions that are not BHCs (e.g., Real Estate Investment Trusts,
insurance companies, etc.) have accounts of cash,
money markets, and other deposits held at BHCs that count on the liability side.}
The value of any financial firm then depends on the payments it gets from its claims on other firms, which might themselves depend on the values of yet other firms, creating rich interdependencies between firms' balance-sheets.

The literature on financial contagion has studied how interbank obligations enable a shock in one part of the system to spread widely, and how this contagion is magnified if banks incur costs upon default.\footnote{See, for instance, Kiyotaki and Moore \citeyearpar{kiyotaki1998credit},
 Eisenberg and Noe
\citeyearpar{eisenbergn2001}, Gai and Kapadia \citeyearpar{gaik2010},
Elliott, Golub, and Jackson \citeyearpar{elliottgj2014},
Acemoglu, Ozdaglar, and Tahbaz-Salehi \citeyearpar{acemogluot2015},
and Jackson and Pernoud \citeyearpar{jacksonp2021b}.  For a recent survey, see
Jackson and Pernoud \citeyearpar{jacksonp2021}.}
 What has received less attention is how these interdependencies can lead to
 multiple equilibria,\footnote{We use the term ``equilibrium''  to keep with the literature,
but here this term only reflects the mutual consistency of banks' values -- so a fixed point in accounting balance sheets -- and not any
 strategic behavior.} with the defaults of some banks becoming
 self-fulfilling. For instance, a Bank A defaulting on its obligation to another Bank B, might lead B to default as well. This can lead to further missed payments that feed back to Bank A, making its initial default self-fulfilling.

Such self-fulfilling defaults on payments are not uncommon in practice, especially in times of distress. For instance,  Fleming and Keane \citeyearpar{fleming2021netting} find that almost 74\% of the settlement failures that occurred in the U.S. Treasury market in March 2020 were effectively ``daisy-chain'' failures, which could have been avoided had all trades been centrally netted.\footnote{Settlement failures were much lower for trades that were centrally cleared (Duffie \citeyearpar{duffie2020still}).}
These defaults, even if only temporary, are costly: delayed in payments can lead counterparty risk to build up, dry up liquidity in the market, restrict or stall investments, and even push some banks towards insolvency.
The existence of multiple equilibria therefore contributes to the fragility of the financial system: pessimistic beliefs or uncertainty about the state of others' balance-sheets can become self-fulfilling and lead banks to stop payments to each other---a type of credit freeze---even when another equilibrium exists in which payments are all made.

In this paper, we characterize the network structures and portfolio returns that lead to self-fulfilling defaults and how to prevent them at minimal cost to the regulator. We consider financial networks in which banks are linked via unsecured debt
contracts.\footnote{The analysis extends to other sorts of contracts as we discuss in the Online Appendix.}
The value of a bank depends on the value of its assets outside of the network,
as well as of its claims on other banks.
If a bank's assets are not enough to cover its liabilities, the bank defaults on
part of its debts.
Importantly, we allow for failure costs that discontinuously depress a bank's balance-sheet
upon default, and reduce the amount repaid to counterparties. We take as given banks' exposures to each other throughout the paper. Our analysis is then most relevant for crises scenarios when banks have little leeway to adjust their positions.

There are two main applications/interpretations of our model. One is when a bank's default involves bankruptcy, and the failure costs are the losses associated with filing for bankruptcy and liquidating assets.\footnote{%
Defaults involve substantial deadweight costs,
including fire sales, early termination of contracts, administrative costs of government bailouts,
and legal costs, among others.
Estimates of bankruptcy recovery rates
can be below 60\%, and even worse in a crisis.
See, for example, Branch \citeyearpar{branch2002},
Acharya, Bharath, and Srinivasan \citeyearpar{acharyabs2007}, and Davydenko, Strebualaev,
and Zhao \citeyearpar{davydenkosz2012}.}
The other major application of the model is to situations in which a bank's default may only be temporary, due to an episode of illiquidity, but can still lead to a sequence of defaults or delayed payments.
Failure costs
then capture costs associated with delayed payments, as outlined above.\footnote{Eisenberg and Noe \citeyearpar{eisenbergn2001} and most of the ensuing literature assume that all payments are cleared \emph{simultaneously}, circumventing this problem. A notable exception is Csoka and Herings \citeyearpar{csokah2018} who show that most decentralized clearing processes yield the lowest equilibrium for bank values, providing additional motivation for our analysis.
Although not explicit in our model, timing on payments can lead to self-fulfilling defaults: e.g., if Bank A must pay Bank B first, and then Bank B can pay Bank C, and then Bank C can pay Bank A, but Bank A cannot make the payments until it receives the payment from Bank C.   We abstract from timing details in the model given the complexity of our analysis, but they offer another interpretation.}
Such defaults are often avoided in practice by short-term borrowing.  However, in situations such as what
occurred in 2008, lenders often lose confidence in the market, which leads to a lack of liquidity for short terms borrowing, and then makes a freeze self-fulfilling.

We make the following contributions.

First, we provide a comprehensive analysis of
the multiplicity of equilibrium bank values.
We show that there exist multiple solutions for bank
values if and only if there exists a certain type of cycle in the network.
Cycles allow costly defaults
to feedback through the financial
network, generating the possibility of self-fulfilling defaults.
There are well-defined `best' and `worst' equilibria, ordered by banks' values.
In the best equilibrium none of the self-fulfilling cycles of defaults occur,\footnote{There can still be some defaults in the best equilibrium, but they are not self-fulfilling. We characterize these in the results below.}
while in the worst equilibrium they all do. There are also
intermediate equilibria that involve some, but not all, of the self-fulfilling cycles defaulting.
We then give necessary and sufficient conditions  for banks'
solvency under both the best and worst equilibria, assuming that a bank cannot make any payments until it
is fully solvent. This applies in practice whenever
insolvencies lead to delays in payments, which can cascade and generate a credit freeze.\footnote{The
insights and main results extend when there are partial payments (a fortiori),
but this case provides the most direct intuition.}
All banks are solvent in the best equilibrium
if and only if their portfolio satisfy an appropriate balance condition. Solvency in other equilibria
is more demanding, and we show how it is precisely characterized by cycles in the network.

Second, we analyze the most cost-efficient way to inject capital into banks so as to avoid
defaults and associated deadweight losses. Injecting capital into a bank has consequences beyond the bank itself:
by paying back its obligations to others, it can bring some of its counterparties closer to
solvency and lower the cost of bailing them out, or even bring them back to
solvency and trigger a repayment cascade.
We show that these indirect-bailout values are critical to understanding  the minimal injections of capital
needed to ensuring systemic solvency.
Building on our analysis of equilibrium multiplicity,
we identify the minimum bailout payments needed to ensure that all banks are solvent in {\sl any} given equilibrium.

The minimum bailouts to ensure systemic solvency in the best equilibrium
do not depend on the specific network
structure, and only require bringing each bank's portfolio into balance.
In contrast, the minimum bailouts ensuring systemic solvency in any other equilibrium
depend on the details of the network structure, as they require injecting enough capital so as
to clear whichever self-fulfilling cycles are defaulting in that equilibrium.
Characterizing the minimum bailouts needed to ensure solvency in a non-best equilibrium is
thus much more complex. In fact, we prove that it is a {\sl strongly NP-hard} problem - which
implies that there are no known practical algorithms for finding approximate solutions, even with relatively small numbers of
banks.
Part of the complexity
comes from the fact that the amount of capital the regulator has to inject in a bank to make it solvent depends on who else is already solvent in the network -- in short, the order of bailouts matters,
and every bank's balance sheet can change with each sequence of bailouts.
Hence the number of bailout policies to consider is of the order of $n!$, where $n$ is the number of banks, and so it is already
in the trillions with just fifteen banks and beyond $10^{18}$ with twenty banks,
and so an exhaustive search for the optimal policy gets infeasible very quickly. 

Despite this complexity, our analysis provides intuitive insights about optimal bailouts.
In particular, indirect bailouts are part of an optimal policy:
when considering the bailout of a particular bank,
instead of injecting capital directly into it, it is often cheaper to inject smaller amounts
into banks that owe that bank money, leveraging those banks' capital. This can be seen
as an explanation of the AIG bailout in 2008, which some argue was an
indirect bailout of Goldman Sachs and others
(Bernard, Capponi and Stiglitz \citeyearpar{bernard2017bail}).\footnote{Roughly,
one could view AIG as a peripheral node as it was mainly selling insurance and
so was a major debtor rather than creditor,
whereas Goldman Sachs is a core dealer and was a major creditor of AIG.}
Building on this intuition, we propose a
simple algorithm that bails out banks in decreasing order of their indirect bailout
value to bailout cost ratio. Even though this algorithm is not always optimal,
we show that it guarantees systemic solvency at a total cost that never exceeds half of the
total overall shortfall.

We then consider some prominent network structures under which there are simpler and intuitive optimal bailout policies
that prevent self-fulfilling defaults and freezes. A key example is a star network in which a core bank is
linked to peripheral banks. Since only the core bank lies on several cycles, finding
the optimal bailout policy is a more tractable problem  -- specifically, it is only
\emph{weakly} NP-hard and collapses to what is known as a Knapsack problem. We show that it is always cheaper to
start by bailing out peripheral banks as opposed to targeting the core bank directly, as it
allows the regulator to leverage peripheral banks' capital buffers.\footnote{Note how considering the network structure yields different insights from the literature on optimal bailout policies when banks are heterogenous. In the latter, it can make sense to bail out stronger banks first as they are closer to solvency (e.g., see Choi \citeyearpar{choi2014heterogeneity}), but this overlooks the indirect value of bailouts.} This again highlights
 the value of indirect bailouts, and we show how some simple bailout
algorithms work well in this setting.

We then discuss more general core-periphery networks,
in which a similar intuition also holds: if peripheral banks are ``small'' relative to core banks,
then it is always optimal to start by bailing out the periphery.
However,
finding the overall
optimal bailout policy remains strongly NP-hard, as core banks are densely connected
and lie on many overlapping cycles.  Thus, although we show that finding optimal bailouts is computationally
challenging, we also show that there are many settings in which parts of the problem can be solved efficiently,
while other parts cannot.  Regardless, the indirect bailout values are important to understand and
can improve over more naive bailout policies.

Finally, we use numerical simulations to compare the performance of various bailout policies in a broader class of core-periphery networks. In line with our analytical results, we show that policies that target peripheral banks first often outperform those that target core banks directly. This is particularly true when peripheral banks have assets of intermediate size. Furthermore, we show that ``naive" bailout policies perform particularly poorly compared to more sophisticated ones when there are large asymmetries among core banks. 

Overall, our results also identify the precise benefit of canceling out cycles of claims---what
has become known as `payment netting' (Kahn and Roberds \citeyearpar{kahn1998payment}, Martin and McAndrews
\citeyearpar{martin2008economic}) and `portfolio
compression' (D'Errico and Roukny \citeyearpar{derricor2019} and Schuldenzucker and
Seuken \citeyearpar{schuldenzucker2019})---and we end the paper with a discussion of such netting techniques.

Our model builds on the literature that followed the interbank lending network
model of Eisenberg and Noe \citeyearpar{eisenbergn2001}.
They introduce the notion of a clearing vector, which specifies mutually consistent repayments on
interbank loans for all banks in the network, and show that it is generically
unique.
Others  have pointed out that the non-negligible  failure costs that banks incur whenever
they are insolvent facilitate the existence of multiple  clearing vectors, and hence of
multiple equilibrium for banks' values
(Rogers and Veraart \citeyearpar{rogers2013failure};
Elliott, Golub, and Jackson \citeyearpar{elliottgj2014})).
This multiplicity comes from the
discontinuous drop in a bank's value at default, which can create self-fulfilling
combinations of defaults.\footnote{This source of equilibrium multiplicity
differs from bank runs \`a la  Diamond and Dybvig \citeyearpar{diamondd1983}.
Here, it stems
from reduced payments due to failure costs that become self-fulfilling, and not from the
optimizing behavior of agents who, anticipating a bank's failure, claim their assets and bring
it to insolvency.}  This multiplicity has not been examined in any detail, and instead
previous studies restrict attention
to equilibrium  repayments that lead to the least number of defaults.

One of the only papers that emphasizes the multiplicity of clearing vectors, and hence of
equilibrium values for banks, is Roukny, Battiston and Stiglitz \citeyearpar{rouknybs2018}.
However, the main focus of their paper differs from ours: they show how equilibrium multiplicity
makes assessing systemic risk harder, as it means some defaults are indeterminate, and propose a
method to measure this source of uncertainty.
We give a new and full characterization of equilibrium multiplicity,
which we then use to analyze and develop optimal bailout policies.

Our analysis of optimal bailouts relates to
Demange \citeyearpar{demange2016}, who characterizes
the optimal cash injection policy in a network of interbank lending.
She defines an institution's threat index as the
marginal impact of an increase in its direct asset holdings on total debt
repayments in the system, assuming the policy does \emph{not}
change the set of defaulting banks. 
Hence a bank's threat index captures its \emph{marginal} social value of liquidity.
In this paper, we instead examine how much of an injection is needed to change and avoid defaults in any equilibrium,
 show how complex that problem is, and offer insights into solving it.\footnote{Since the first writing of this paper, others have also analyzed complexity of bailouts (Egressy and Wattenhofer \citeyearpar{egressy2021bailouts}, Klages-Mundt and Minca \citeyearpar{klages2022optimal}). They focus on the best equilibrium and consider other objective functions, more in line with our discussion in Section \ref{budget} than with our main analysis.}


Several papers highlight various distortions induced by bailouts in financial networks.
Erol \citeyearpar{erol2019} shows that public bailouts affect banks' choice of counterparty,
and hence the equilibrium structure of the financial network. Leitner \citeyearpar{leitner2005} and 
Kanik \citeyearpar{kanik2018} study how linkages
between banks can incentivize private sector bailouts, whereby solvent banks bail out insolvent ones,
and how this depends on the network structure.
Bernard, Capponi and Stiglitz \citeyearpar{bernard2017bail} analyze the interplay between
public bailouts and private bail-ins. Capponi, Corell, and
Stiglitz \citeyearpar{capponi2020optimal} investigate how
debt-financed bailouts increase the risk premia on sovereign debt, and hence depress
the balance sheets of the very banks the regulator was trying to save, generating a doom loop.
Our paper is orthogonal to this literature, as we examine the structure and complexity of bailouts \emph{ex-post} rather than
their external effects on decisions that are made by banks \emph{ex-ante}.

Finally, the literature on unique implementation in games with strategic complementarities (e.g., Segal \citeyearpar{segal2003coordination} and Winter \citeyearpar{winter2004incentives}) is worth noting. Following Eisenberg and Noe \citeyearpar{eisenbergn2001}, we do not model interbank repayments as a game but simply presume that banks repay as much of their debts as they can. An equivalent formulation would allow banks to choose how much to repay each other, with payoffs such that banks prefer paying debts when they can. The induced game would feature strategic complementarities and be prone to multiple equilibria.\footnote{Bank $i$ can pay back (weakly) more of its liabilities if others do so as well.}  The optimal bailouts preventing defaults in the worst equilibrium are then the minimal transfers ensuring banks coordinate on the best equilibrium. This is in the spirit of that strand of the literature on unique implementation, which looks for mechanisms that coordinate agents on the best equilibrium in the eyes of the designer.


\section{A Model of Financial Interdependencies}

\subsection{Financial Institutions and their Portfolios}

Consider a set $N= \{0, 1,\ldots,n\}$ of organizations involved in the network.
We treat $ \{1,\ldots, n\}$ as the financial organizations, or ``banks" for simplicity in terminology.
These should be interpreted as a broad variety of financial organizations, including banks,
venture capital funds, broker-dealers, central counterparties (CCPs), insurance companies, and
many other sorts of shadow banks that have substantial financial exposures on both sides of their balance sheets.
These are organizations that can issue as well as hold debt.

We lump all other actors into node $0$ as these are entities that either hold debt in the financial organizations (for instance
private investors and depositors), or borrow from the financial organizations
(for instance, most private and public companies), but not both.
Their balance sheets may be of interest as well, as the defaults on mortgages or other loans could be
important triggers of a financial crisis.  The important part about
the actors in node $0$ is that, although they
may be the initial trigger and/or the ultimate bearers of the costs of a financial crisis, they are not  the dominoes, becoming insolvent and defaulting on payments as a result of defaults on their assets.
In aggregate, it may appear that there is debt going both in and out of node 0, but none of the individual
private investors that comprise node 0 have debt coming both in and out.%
\footnote{Of course, this is an approximation and there is a spectrum that involves a lot of gray area.
For instance, Harvard University invests tens of billions of dollars, including making large loans.  At the same time it borrows money and has issued debt of more
than five billion dollars.  It is far from being a bank, but still has incoming and outgoing debt and other obligations.
This is true of many large businesses, and all those that could become dominoes
should be included in $\{1,\ldots,n\}$.
It is not so important for us to draw an arbitrary line through this grey area to make our
points.    Nonetheless, this is something a regulator does have to take a stand on
when trying to address systemic risk, and in practice may even be dictated by jurisdictional rules.  }

Bank portfolios are composed of investments in assets outside the system as well as financial contracts within the system. Investments in primitive assets involve some initial investment of capital and then pay off some cash flows over time, often randomly; e.g., government securities, asset backed securities, corporate and private loans, mortgages, etc.
This is part of a bank's capital.
We let $p_i$ denote the current total market value of $i$'s investments in those assets and
$\mathbf{p}=(p_i)_i$ denote the associated vector. (Analogous bold notation is used for all
vectors and matrices.)

The book values of banks in the network are based not only on the capital in these outside investments, but
also on assets from and liabilities to others in the financial system.
In this paper, we restrict attention to interbank debt contracts, but the model extends to allow for
other types of financial contracts between banks.\footnote{See Appendix \ref{sec:equity} of the online
material for an extension of the model and results when banks also hold equity claims on each other.}
A debt contract between a creditor $i$ and a debtor $j$ is characterized by its face
value $D_{ij}$. As a bank cannot have a debt on itself, we set $D_{ii}=0$ for
all $i$. Let $\mathbf{D}$ the matrix whose $(i,j)$-th entry is equal to $D_{ij}$, and denote a
bank's total nominal debt assets and liabilities by, respectively, $D_i^{A}\equiv \sum_j D_{ij} $
and $D_i^{L}\equiv \sum_j D_{ji}$.

\subsubsection{The Weighted Directed Network}

The financial network generated by interbank lending contracts is thus represented by a weighted directed graph on $N$, where a directed edge pointing from bank $i$ to bank $j$ means $i$ has a debt liability toward $j$ with a weight of $D_{ji}$, so edges point from the debtor to the creditor.

A {\sl directed path} in the network from $i$ to $j$ is a sequence of banks $i_0, \dots i_K$, for
some $K\geq 2$ such that:
$i_0=i$ and $i_K=j$ and $D_{i_{\ell+1}i_{\ell}}>0$ for each $\ell<K$.  Thus a directed path is such that $i$ owes a debt to some bank, which owes a debt to another bank, and so forth, until $j$ is reached.

A {\sl dependency cycle}, or {\sl cycle} for short, is a directed cycle in the network which is a sequence of banks $i_0, \dots i_K$, for
some $K\geq 2$ such that:
$i_0=i_K$ and $D_{i_{\ell+1}i_{\ell}}>0$ for each $\ell<K$.  A directed cycle is {\sl simple} if $i_0$ is the only bank repeated in the sequence.

Given that node 0 is comprised of entities that cannot be involved in cycles,  to avoid confusion with the network of
debts, we set $D_{i0}=0$ for all $i$, and any debts owed from outside of the network of banks are instead recorded in $p_i$.  Outsiders can still default on a payment to a bank, but that is captured in a lower value of $p_i$.

\subsection{Defaults and Equilibrium Values of Banks}

A main object of interest in this paper is a bank's book value $V_i$. Following  Eisenberg and Noe \citeyearpar{eisenbergn2001}, we focus on bank values ex post, taking as given their exposures to each other and realized portfolios. 
We first need to introduce defaults and their associated costs, before characterizing bank values.

If the value of bank $i$'s assets falls below the value of its
liabilities, the bank  is said to \emph{fail} and incurs \emph{failure costs}
$\beta_i(\mathbf{V},\mathbf{p})\geq 0$. 
These costs capture the fact that the value of a bank's
balance sheet can be discontinuously depressed upon insolvency,
for instance due to direct costs of bankruptcy: legal and auditors' fees,
fire sales, premature withdrawals, or other losses associated with halted or decreased operations.

In our setting, $\beta_i$ can also capture some indirect costs: it is not necessary that the bank declares bankruptcy,
but simply that its insolvency causes it to renegotiate its contracts or delay payments, leading to a freeze and
imposing costs on it and its creditors.
Such failure costs can depend on the degree to which $i$ and others are
insolvent as well as the value
 of their portfolios.
Hence we allow these costs to depend on the vector of bank values $\mathbf{V}$ and returns
on outside investments $\mathbf{p}$.\footnote{We have reduced the portfolios to only track their total
value, but in practice the failure costs incurred could depend on detailed information about the
composition of the bank's portfolio as well as the portfolios of others.
 }

With the possibility of default,  the realized debt payment from a bank $j$ to one of its creditors $i$
 depends on the value of $V_j$, and thus its solution depends
ultimately on the full vector $\mathbf{V}$.
To make these interdependencies explicit,
let $d_{ij}(\mathbf{V})$ denote the amount
of debt that bank $j$ actually pays back to $i$.

There are two regimes. If bank $j$ remains solvent, it
can repay its creditors in full, and then for all $i$
\[
d_{ij}(\mathbf{V}) =   D_{ij}.
\]
If instead $j$ defaults,
then debt holders become the residual claimants, and are rationed proportionally to their claim on $j$:
\begin{equation}
\label{debtvalue}
d_{ij}(\mathbf{V}) =   \frac{D_{ij}}{\sum_h D_{hj}} \max\left(p_j+ d_j^A(\mathbf{V})
- \beta_j(\mathbf{V},\mathbf{p}),0\right),
\end{equation}
with $d_j^A(\mathbf{V})=\sum_i d_{ji}(\mathbf{V})$.

It is useful to introduce notation for a bank's realized failure
costs, $b_i\left(\mathbf{V},\mathbf{p}\right)$, which are 0 if the bank remains solvent
and equal to $ \beta_i\left(\mathbf{V},\mathbf{p}\right)$ if it
defaults:\footnote{This allows us to use the same notation for solvent and insolvent banks,
and makes equilibrium bank values easier to write down.}
\begin{equation}
\label{bankrupt}
  b_i\left(\mathbf{V},\mathbf{p}\right) =
  \begin{cases}
   0 & \text{ if  }p_i  +d_i^A(\mathbf{V})  \geq D_i^L \\
   \beta_i\left(\mathbf{V},\mathbf{p}\right) & \text{ if  }p_i  +d_i^A(\mathbf{V})  <D_i^L.
  \end{cases}
\end{equation}

We add conditions on failure costs to avoid the possibility that these costs per se, and not the reduced payments they imply, induce multiple equilibria for bank values.
The key assumption is that failure costs cannot increase faster than the value of a bank's assets.
In particular, we assume that
$d_i^A(\mathbf{V}) -\beta_i(\mathbf{V},\mathbf{p})$ is nondecreasing in $\mathbf{V}$.

%

Note also that we have carefully written failure costs
$b_i\left(\mathbf{V},\mathbf{p}\right) $ as a function of
how a bank's assets $p_i  +d_i^A(\mathbf{V})$
compare to its liabilities $D_i^L$, instead of
as a function of $V_i$ directly.   This avoids having defaults or freezes driven solely by the anticipation of
failure costs, even when a bank
has more than enough assets, even cash on hand, to cover its liabilities.
Such a self-fulfilling default would go beyond a bank run, since it would not be due to the bank
not having enough cash on hand to pay its debts.
It would instead be due to self-fulfilling failure costs with no interaction with the financial network
or portfolio values.   We rule this out as it seems of no practical interest.
However, as noted above, we do allow realized failure costs $ \beta_i\left(\mathbf{V},\mathbf{p}\right) $
to depend on the value of other banks; e.g., a bank can incur higher costs upon default if
others are defaulting as well, because of fire sales.
Such a dependency is important as it can worsen contagion, and we include it in
our analysis.

A canonical example of admissible failure costs corresponds to the case where
\begin{equation}\label{canon}
\beta_i(\mathbf{V},\mathbf{p})=b+a\left[ p_i
+d_i^A(\mathbf{V})\right]
\end{equation}
with $b\geq 0$ and $a\in [0,1]$.\footnote{If $b=0$, these failure costs are similar to the ones considered in
Rogers and Veraart \citeyearpar{rogers2013failure}, who also observe that costs of default {can} generate equilibrium multiplicity. They, however, do not study which network structures generate multiple equilibria nor how to prevent such multiplicity, and instead restrict attention to the best equilibrium to study rescue consortia (see the discussion in Kanik \citeyearpar{kanik2018}).}
In that case, failure costs are composed of some fixed amount (e.g., legal costs),
as well as some share of the value of the bank's assets. This is a reasonable
assumption if, for instance, the bank only recovers some fraction
of its assets upon sale
(e.g., due to a markdown on a fire sale of its assets) or has a portion of its legal costs
that scales with the size of the enterprise.\footnote{Several papers estimate marginal bankruptcy costs
being around 20\% or 30\% of the value of a bank's assets (see for instance Davydenko et al. \citeyearpar{davydenkosz2012}),
suggesting $a$ should be in that range. As mentioned above, these can be even larger, Branch \citeyearpar{branch2002},
Acharya, Bharath, and Srinivasan \citeyearpar{acharyabs2007}, particularly in times of a financial crisis.
Fixed costs associated with bankruptcies are harder to
estimate, but are presumably strictly positive given the structure of legal and accounting costs, as
well as the fact
that marginal cost estimates are much lower than overall costs.}

It can be that failure costs exceed the value of the defaulting bank's assets;
e.g., if $b$ is large enough in the above example. These excess costs should be interpreted
as real costs, for instance, debts or legal costs that are never paid, capital or labor that
are idled,  etc., which can be incurred by the bank itself if it does not act under limited liability,
or by the government or agents outside of the
network (so node $0$ in our framework).
What matters for our analysis is not whether failure costs can exceed the bank's
assets, but that they crowd out some of the debt repayments to creditors.

Failure costs are imposed on banks' balance sheets directly,
and so the book value of a bank $i$ is
\begin{equation}
\label{eq-bookvalue-bankruptcy-i}
V_i=p_i +   d_i^A(\mathbf{V})-D^L_i  - b_i(\mathbf{V},\mathbf{p}),
\end{equation}
where $b_i(\mathbf{V},\mathbf{p})$ is defined by (\ref{bankrupt})
and $d_i^A(\mathbf{V})$ by (\ref{debtvalue}) whenever there are some
defaults.
In matrix notation:
\begin{equation}
\label{eq-bookvalue-bankruptcy}
\mathbf{V}=  \mathbf{p} + \mathbf{d}^A(\mathbf{V})  - \mathbf{D}^L - \mathbf{b}(\mathbf{V},\mathbf{p}).
\end{equation}

A vector of bank values $\mathbf{V}$ is an \emph{equilibrium} if it is a solution to
equation (\ref{eq-bookvalue-bankruptcy}).
Note that these are banks' equilibrium values ({ex post}), at the time of settlement,
once returns and defaults are realized. Any shock to outside investments is already realized, observed, and encoded into $ \mathbf{p}$. Equilibrium values can be negative
if a bank's liabilities exceed the value of its assets. These can then be interpreted as
``hypothetical'' values and not equity values.  If a bank $i$ is solvent, then its equity
value coincides with $V_i$.
However, if its ``hypothetical'' value $V_i$ is negative then $i$ defaults, and
its equity value is zero.
The extra negative value means that a bank's assets are not enough to cover its liabilities,
and hence that some debt payments are not made in full.
Coupled with failure costs, there are deadweight losses in the economy.

\section{Multiplicity of Bank Values and Self-Fulfilling Defaults}

Although the possibility of multiple equilibria for bank values in financial networks is well-known, the conditions under which they exist and their implications are not. In this section we characterize when there exists a multiplicity. We also derive necessary and sufficient conditions on portfolio values for all banks to be solvent, which depend on the equilibrium being considered.

All of the definitions that follow are relative to some
specification of
$\mathbf{p,D}$, and we omit its mention.

\subsection{The Multiplicity of Equilibrium Values} \label{sec:lattice}

Because the value of a bank is weakly increasing in others' values,
there always exists a solution to equation (\ref{eq-bookvalue-bankruptcy}).
Furthermore, there can exist multiple solutions given the interdependence
in values, and in fact, the set of equilibrium values forms a complete lattice.\footnote{This can be seen by an application of
Tarski's fixed point theorem, since banks' values depend monotonically on each other.
They are bounded above by the
maximum values of banks' assets $\textbf{p}+\textbf{D}^A$.
 }
Thus, there exists a ``best''
as well as a ``worst''
equilibrium, in which bank values hit an overall maximum and minimum,
respectively.  The set of defaulting banks is hence the largest in the worst equilibrium,
and the smallest in the best.

The following algorithm finds the best equilibrium. Start from bank values $\textbf{V}^0$ that are at least as high as the best equilibrium values (e.g.,
$\textbf{V}^0 =\textbf{p}+\textbf{D}^A$)
and then compute
$$\textbf{V}^1 = \textbf{p}+\textbf{d}^A(\textbf{V}^0)
- \textbf{D}^L - \textbf{b}(\textbf{V}^0,\textbf{p}).$$
If any values are negative, the associated banks default and the values are computed again accordingly.
Iterating this process yields the best equilibrium.
The worst equilibrium can be found using a similar algorithm, but starting from
values that must lie below the worst equilibrium values (e.g.,
$ \textbf{p} - \textbf{D}^L - \overline{\textbf{b}}$, where $\overline{\textbf{b}}$ is a
cap on how large failure costs can be).

\begin{figure}[!h]
\begin{center}
\begin{tikzpicture}[scale=1.2]
\definecolor{afblue}{rgb}{0.36, 0.54, 0.66};
\foreach \Point/\PointLabel in {(0,0)/1, (6,0)/3}
\draw[ fill=afblue!40] \Point circle (0.35) node {$\PointLabel$};
\draw[fill=afblue!40] (3,0) circle (0.35) node {$2$};
\draw[->, thick] (0.4,0.4) to [out=40,in=135] node[midway,fill=white]  {$D_{21}=1$} (2.6,0.4);
\draw[<-, thick] (0.4,-0.4) to [out=-40,in=-135] node [ midway,fill=white]  {$D_{12}=1$} (2.6,-0.4);
\draw[<-, thick] (3.4,0.4) to [out=40,in=135] node [ midway,fill=white]  {$D_{23}=1$} (5.6,0.4);
\draw[->, thick] (3.4,-0.4) to [out=-40,in=-135] node [ midway,fill=white]  {$D_{32}=1$} (5.6,-0.4);
  \end{tikzpicture}
  \end{center}
  \captionsetup{singlelinecheck=off}
  \caption[]{An Example of Equilibrium Multiplicity.  Arrows point in the direction that debt is owed,
  that is from debtors to creditors.}
  \label{fig:algorithm}
\end{figure}

Figure \ref{fig:algorithm} provides a simple example, for which we compute bank values using these two algorithms.\footnote{See Section \ref{sec:multiple_ex} of the Appendix for an example with three different equilibria for bank values.}
Suppose $p_1=1$, $p_2=p_3=0$, and $\beta_i(\textbf{V}, \textbf{p})=0.5[p_i +
d_i^A(\textbf{V})]$ for $i=1,2,3$. So a bank loses half of the value of its assets upon default.
Let us first derive the best equilibrium for bank values using the above algorithm. We initiate at
$\textbf{V}^0 =\textbf{p}+\textbf{D}^A=(2,2,1)$. Since $\textbf{V}^0\geq\textbf{0}$, no
failure costs are incurred $ \textbf{b}(\textbf{V}^0,\textbf{p})=\textbf{0}$ and all
debts are repaid $\textbf{d}^A(\textbf{V}^0)=\textbf{D}^A$. Then $\textbf{V}^1=(1,0,0)$.
Since $\textbf{V}^1\geq\textbf{0}$ the algorithm stops, and bank values in the best equilibrium are
$\overline{\textbf{V}}=(1,0,0)$. \\ We now derive the worst equilibrium for bank values.
Note that failure costs cannot be greater than $\overline{\textbf{b}}=0.5[\textbf{p}+\textbf{D}^A]$. We hence initiate the algorithm at $\textbf{V}^0 =\textbf{p}-\textbf{D}^L-\overline{\textbf{b}}=(-1,-3,-1.5)$. Since $p_1+d_1^A(\textbf{V}^0) \geq p_1= D_1^L$, it has to be that bank 1 does not default: $b_1(\textbf{V}^0, \textbf{p}) = 0$ and $\textbf{d}_{21}(\textbf{V}^0)=D_{21}$.  Debt repayments for bank 2 and 3 solve
  \[  \begin{cases} d_{23}(\textbf{V}^0) =\frac{D_{23}}{D_3^L}0.5[p_3+D_{31}+d_{32}(\textbf{V}^0)] = 0.5 d_{32}(\textbf{V}^0) \\
  d_{32}(\textbf{V}^0) =\frac{D_{32}}{D_2^L}0.5[p_2+D_{21}+d_{23}(\textbf{V}^0)] = 0.25[1+d_{23}(\textbf{V}^0) ]
  \end{cases} \iff  \begin{cases} d_{23}(\textbf{V}^0)=\frac{1}{7}\\ d_{32}(\textbf{V}^0) =  \frac{2}{7}\end{cases}.\]
Hence,  $\textbf{d}^A(\textbf{V}^0)= \left(\frac{2}{7},  \frac{8}{7}, \frac{2}{7}\right)$, and $\textbf{V}^1 = \textbf{p}+\textbf{d}^A(\textbf{V}^0)- \textbf{D}^L - \textbf{b}(\textbf{V}^0,\textbf{p})= \left(\frac{2}{7}, - \frac{10}{7}, -\frac{6}{7}\right)$. Since no new solvencies are induced -- indeed $p_i+d_i^A(\textbf{V}^1)<D_i^L$ for $i=2,3$ -- the algorithm stops, and bank values in the worst equilibrium are $\underline{\textbf{V}}=  \left(\frac{2}{7}, - \frac{10}{7}, -\frac{6}{7}\right)$.

When there are multiple equilibria, any equilibrium other than the best equilibrium
must involve cycles of defaults that could have been avoided: these are defaults that are
triggered either by pessimistic beliefs about the balance-sheets of others---valuing their debt at
a low value---which become self-fulfilling, or by a liquidity freeze in which debts are not paid in and thus not paid out.
These avoidable defaults are essentially coordination failures:
cycles of banks could have written-off (some) of their counterparties' debt so as to avoid at least
 some of the defaults and associated costs.
  Given that financial markets are prone to runs and freezes, understanding when such self-fulfilling
  cascades exist is of practical importance.

The following proposition highlights how equilibrium multiplicity depends on the
presence of cycles of liabilities, combined with failure costs.

\begin{proposition}\label{uniqueV}
 For each $i$, let the failure costs $\beta_i(\cdot, \mathbf{p})$ depend only on $i$'s
 value $V_i$ and the values $V_j$ of banks $j$ on which $i$ has a (potentially indirect)
 debt claim, 
and be a contraction as a function of $V_i$ (e.g., the canonical failure costs in (\ref{canon})).
\begin{itemize}
\item[(i)] If there is no dependency cycle, then the worst and best (and thus all) equilibria coincide.
\item[(ii)] Conversely, if there is a dependency cycle, then there exist failure costs
(satisfying the above conditions)
and values of bank investment portfolios $\mathbf{p}$ such that the
best and worst equilibria differ.
\item[(iii)] Any equilibrium that differs from the best equilibrium consists of
the defaults in the best equilibrium plus the defaults of all banks that lie on some set of dependency cycles.
Any other banks defaulting in this equilibrium but not in the best equilibrium
lie on outpointing paths from the original defaulting banks and the newly
defaulting dependency cycles.\footnote{It is possible that the new defaults on an
additional dependency cycle leads banks that defaulted in the best equilibrium to
payout even less than in the best equilibrium, which can lead to additional defaults
on rays outwards from the original banks, not just the newly defaulting banks.}
\end{itemize}
\end{proposition}

We are not the first to notice that cycles are necessary to generate multiple equilibria,
as they are required for self-fulfilling feedbacks
(e.g., see Elliott et al. \citeyearpar{elliottgj2014}, Roukny et al. \citeyearpar{rouknybs2018}),
and a similar intuition is at play here and underlies part (i).
Proposition \ref{uniqueV}
however goes beyond that as it proves that cycles are also sufficient (part (ii)),\footnote{They are sufficient in the sense that they always enable equilibrium multiplicity for some realization of portfolios and failure costs.}
and it  highlights the general structure of the set of equilibria (part iii).
This last observation is
particularly useful, as even though there is a lattice of equilibria, figuring
out what those equilibria are could be quite complex.
This result ensures that one can find all equilibria by examining combinations of
cycles, which can greatly simplify the calculations.

Although proving part (i) is straightforward,
proving part (ii) requires finding returns $\mathbf{p}$ and
failure costs $\mathbf{\beta}_i(\cdot, \mathbf{p})$ that generate multiple
equilibria.  The specific  failure costs we use in the proof are
$\beta_i(\mathbf{V},\mathbf{p}) = a(p_i  +d_i^A(\mathbf{V}))$ for $a$ above some threshold;
that is, a bank loses a fraction of its assets when defaulting.
Hence, we do not need to construct costs that depend on other banks'
values to generate multiplicity: it is enough for them to only depend on the value of the
defaulting bank's assets.
It is also not necessary for banks to lose the entirety of their assets ($a=1$) for the
multiplicity to arise, though we assume they do in our analysis of minimum bailouts
for the sake of transparency.

Part (iii) is intuitive, but is important to state as it is very helpful in
calculating equilibria.\footnote{Even in problems with complementarities and a complete lattice of equilibria,
finding all equilibria can be very challenging (see Echenique \citeyearpar{echenique2007finding}
for an algorithm).}  Essentially, beyond the best equilibrium,
all other equilibria involve
self-fulfilling cycles of defaults;
and this can focus the search for equilibria.
Those cycles can also lead to additional casualties that are on
outward paths from an originally defaulting bank or an additional cycle, but the
equilibria must be based on some
dependency cycles.   Not just any combination of dependency cycles will work
(some banks might be strong enough to never default), but each equilibrium must
differ from other equilibria by the inclusion/exclusion of
at least one cycle. 

With the restriction that failure costs can only depend on a bank's direct and indirect neighbors,
the feedback that leads to multiple equilibria can only occur through a dependency cycle
 (Proposition \ref{uniqueV} (i)).
If instead, the costs incurred by bank $i$ upon default are larger if $j$ defaults as well,
even though $i$ has no network path to $j$, then failure costs themselves can generate ``indirect''
cycles.  For example, if fire sales change bank values, then $i$'s value could depend on whether $j$ becomes insolvent even if $i$ is
not path-connected to $j$ in the network of debts.
It is then possible to have multiple equilibria for bank values even
in the absence of any cycle in the network of debt.
The following example illustrates this.
\begin{figure}[!h]
\begin{center}
\begin{tikzpicture}[scale=0.8]
\definecolor{afblue}{rgb}{0.36, 0.54, 0.66};
\foreach \Point/\PointLabel in {(0,0)/2, (5,0)/1, (10,0)/0}
\draw[fill=afblue!40] \Point circle (0.35) node {$\PointLabel$};
\draw[->, thick] (0.7,0) to node [ above]  {\small $D_{12}= 1$} (4.3,0);
\draw[->, thick] (5.7,0) to node [ above]  {\small $D_{01}= 1+\varepsilon$} (9.3,0);
  \end{tikzpicture}
  \end{center}
  \captionsetup{singlelinecheck=off}
  \caption[]{An Example of Indirect Cycles.
  }
\end{figure}

Suppose Bank 2 and Bank 1's outside investments yield
  $p_2=1-\varepsilon$ and $p_1=1$, respectively. Since Bank 2 has no other assets,
  it must default in any equilibrium. Bank 1's outside investments are not quite enough
  to pay back its liabilities, but a small repayment from Bank 2 would
  be enough to ensure its solvency. Suppose Bank 2's failure costs are small if Bank 1 is
  solvent, $\beta_2(\mathbf{V},\mathbf{p}) = \varepsilon\mathbbm{1}\{V_1\geq 0\}$,
  but large if it is not, $\beta_2(\mathbf{V},\mathbf{p}) = \mathbbm{1}\{V_1< 0\}$.
  There exist two equilibria despite the absence of cycles in the network.
  In the first equilibrium, Bank 1 remains solvent and Bank 2 incurs a small cost, and
  thus repays $D_{12}(\mathbf{V}) = 1-2\varepsilon$ to Bank 1 (which then indeed has
  enough assets to be solvent). In the second equilibrium, Bank 1 defaults and
  Bank 2 incurs a large cost and repays nothing to Bank 1.

One can extend Proposition \ref{uniqueV} to take into account such fire-sale cycles
that arise under more general failure costs. To do so, the network must be augmented  to account for all the ways a bank's value can influence another's.  That is, it must feature an edge from bank $i$ to $j$ whenever $j$'s failure costs $\beta_j(\cdot, \mathbf{p})$ vary with $i$'s value $V_i$, even if $i$ has no obligation to $j$. If that augmented network has no cycles, then there exists a unique equilibrium for bank values. Otherwise, fire sales can generate equilibrium multiplicity even in the absence of cycles of liabilities, as illustrated above.

For the remainder of the paper, we analyze the extreme but illuminating case in which even if a bank has some
money coming in it cannot use that money to pay some of its debt until it is fully solvent.
In particular, we assume ``Full Bankruptcy Costs:''
$$\beta_i(\textbf{V}, \mathbf{p})=p_i + d_i^A(\textbf{V}) {\rm \ for \ each \ bank \ } i.$$

Under this specification for failure costs, the worst equilibrium for bank values arises whenever we take the timing of payments seriously. Indeed, the algorithm that finds the worst equilibrium starts by assuming everyone defaults, which here means that no one makes any debt payments. The only solvent banks are then those that are able to make their payments out even without receiving any payments in. This first wave of payments might enable other banks to meet their obligations, etc, and iterating on this yields the worst equilibrium for bank values. Hence this assumption applies in practice when insolvency leads to delays in payments, which can lead to cascading delays and cycles, and thus at least to a temporary freeze.

These failure costs make solvency
more demanding: the set of defaulting banks can be strictly larger under this rule than when
partial repayments are allowed.  Nonetheless, this case provides the basic intuitions and insights without cluttering the calculations with partial payments.
Some of our results do not rely on this assumption, and in particular results
from Section \ref{bailouts} on the complexity of bailouts, which only
require failure costs to be strictly positive. Our characterizations of systemic
solvency and of the optimal bailout policy in some specific networks do leverage
this assumption, and are more complicated to state without it, even though the underlying
intuitions remain.

\subsection{A Characterization of Systemic Solvency}

Proposition \ref{uniqueV} shows that cycles of debts are vital to the existence of multiple
equilibria, and that such multiplicity means the defaults of some banks can be self-fulfilling.
In this section, we investigate in more details which banks default, and how it depends on the
equilibrium being considered. We give necessary and sufficient conditions on portfolio values
for \emph{all} banks to be solvent, which we call \emph{systemic solvency}.

\subsubsection{Balance Conditions}

To characterize solvency, and minimal bailouts,
it is useful to define the following balance conditions.

We say that a bank $i$ is {\sl weakly balanced} if
\begin{equation*}
p_i+D_i^{A} \geq  D_i^{L},
\end{equation*}
and that the network is {\sl weakly balanced} if this holds for all $i$.

Weak balance requires that a bank's assets are enough to cover its
debt liabilities, presuming its incoming debt assets  are all fully valued.
A network being weakly balanced is sufficient for all banks to be solvent in the best equilibrium.
This follows since, if all banks but $i$ honor their debt contracts, then $i$ can also pay back its debt fully in a weakly balanced network. Essentially, all debts can be canceled out,  irrespective of the network structure.
Things are different in the worst equilibrium, as we know from Proposition \ref{uniqueV}.

We say that a bank $i$ is
{\sl exactly balanced} if
\begin{equation*}
p_i+D_i^{A} =  D_i^{L},
\end{equation*}
and that the network is {\sl exactly balanced} if this holds for all $i$.
An exactly balanced bank has no capital buffer: its assets, if fully valued, are exactly enough to
cover its liabilities.
Exact balance is a much stronger condition than weak balance, but is very useful as a benchmark condition in characterizing
optimal bailouts, as we shall see.

We say that a bank $i$ is
{\sl critically balanced} if it is weakly balanced
and
for each $j$ for which $D_{ij}>0$,
\begin{equation*}
p_i+D_i^{A} - D_{ij} <  D_i^{L}.
\end{equation*}
The network is {\sl critically balanced} if this holds for all $i$.

Critical balance is a special case of weak balance which includes exact balance as a special case, but is not as restrictive and is useful as a benchmark
condition in characterizing optimal bailouts.
It implies that not receiving any one of its incoming debt payments is enough to make a bank insolvent.


\subsubsection{Characterizing Systemic Solvency}

When there exist cycles of debt, by Proposition \ref{uniqueV},  the best and worst equilibria can differ.
We now fully characterize when these cycles fail to clear.

Determining whether cycles clear at a given vector of portfolio
values $\mathbf{p}$ involves an iterative definition of solvency, since a bank being solvent can affect
the solvency of other banks.

We say that a bank $i$ is {\sl unilaterally solvent }
if $p_i \geq D_i^L$.   This means that regardless of whether any
of the other banks pay
the debts that they owe to $i$, $i$ is still able to cover its liabilities.

A set of banks $S$ is {\sl iteratively strongly solvent}
if it is the union of sets $S=S_1\cup \cdots S_K$, such that
 banks in $S_1$ are
unilaterally solvent; and then banks in any $S_k$ for $k\in \{2,\ldots, K\}$ are solvent whenever they receive the debts from all banks in sets
$S_1,\ldots S_{k-1}$:
\[
p_i + \sum_{j\in S_1\cup \cdots S_{k-1}} D_{ij} \, \geq \, D_i^L.
\]

Note that if $N$ is iteratively strongly solvent, then all banks are
solvent in the worst equilibrium.
Proposition \ref{solvency} provides
weaker conditions that are necessary and sufficient for systemic solvency.
This then provides a base to understand optimal bailout policies.

\begin{proposition}
\label{solvency}
\

All banks are solvent in the best equilibrium if and only if the network is
weakly balanced.

All banks are solvent in the worst equilibrium if and only
if the network is weakly balanced and there exists an
iteratively strongly solvent set that intersects each directed (simple)
cycle.\footnote{A simple cycle is such that the only repeated bank is the starting/ending bank.
Note that something intersects each cycle {\sl if and only if} it intersects each simple cycle, and so
the statement is correct whether or not it is restricted to simple cycles. }
\end{proposition}

An implication of Proposition \ref{solvency} is that in a weakly balanced network, if
there is an iteratively strongly solvent set that intersects each cycle, then that implies that the whole set of banks is iteratively strongly solvent.   This is the crux of the proof.

The proposition is less obvious than it appears since an insolvent bank can lie on several cycles at once, and could need all of its incoming debts to be paid before it can pay any out.  Solvent banks on different cycles could lie at different distances from an insolvent bank, and showing that each bank eventually gets all of its incoming debts paid before paying any of its outgoing debts is subtle. The proof is based on how directed simple cycles must work in a weakly balanced network and appears in the appendix.

Proposition \ref{solvency} is useful for at least two reasons.
First, it highlights the structure of the set of equilibria, which is very useful
in our analysis of optimal bailouts.
Second, it gives necessary and sufficient
conditions to find the worst (and other) equilibria that are easier to check  than preexisting algorithms.
Example \ref{fig:ISSS} illustrates this.
\begin{figure}[!h]
\centering
\includegraphics[width=0.85\textwidth]{ISSS_example.tikz}
\caption{\small Arrows point in the direction that debt is owed.
Let $p_1=p_2=p_3=0.2$, $p_4=p_5=1$, and $p_6=p_7=p_8=p_9=0.5$. The only iteratively strongly
solvent set is $\{\{5\}, \{4\}, \{3\}, \{2\}, \{1\}\}$.}
\label{fig:ISSS}
\end{figure}

As all banks have as much debt in as out, they are weakly balanced and
hence all solvent in the best equilibrium.
Identifying which banks default in the worst
equilibrium requires checking the iterative condition.
Bank 5 is the only unilaterally solvent
bank. Its solvency ensures the solvency of Bank 4, and so there is at least an iteratively
strongly solvent set intersecting the left and middle cycles. However, Bank 4 paying back its
debt is not enough to make Bank 6 solvent, and so we can stop the iteration there: no iteratively
strongly solvent set intersects the right cycle, and all banks on that cycle must default in the
worst equilibrium. Interestingly, this is faster than checking that the \emph{entire} set of banks
forms an iteratively strongly solvent set, as we can stop checking banks' solvencies once we have
reached key banks that lie at the intersection of multiple cycles (here Banks 4 and 6). It is also
faster than algorithms that have been developed to find Nash equilibria of
games with strategic complements.\footnote{Equilibrium values for banks correspond to Nash
equilibrium outcomes of an auxiliary game in which banks choose whether to be solvent ($s_i=1$)
or not, and best responses are $s_i=1 \iff p_i +\sum_j D_{ij}s_j\geq D_i^L$.} The worst
equilibrium is usually found by iterating on best-responses starting from the lowest
strategies for all agents (see Echenique \citeyearpar{echenique2007finding}, and references
therein), which in our setting boils down to checking whether the whole set of banks is
iteratively strongly solvent.

Proposition \ref{solvency} also implies that if both conditions are satisfied, then there is a unique equilibrium.
Conversely, if the network is weakly balanced but there is no iteratively strongly solvent set
intersecting every cycle, then there are necessarily multiple equilibria.
Thus, in a weakly balanced network, there is a unique equilibrium {\sl if and only if} there exists
an iteratively strongly solvent set that intersects each directed (simple) cycle.

Clearly, without some unilaterally solvent banks, having an iteratively solvent set is precluded,
and all banks must default in the worst equilibrium. This is not directly implied by
Proposition \ref{solvency}, but still follows from the reasoning behind it.
To see why this is true, recall the algorithm introduced in Section \ref{sec:lattice} to
find the worst equilibrium. The algorithm initiates at values $\mathbf{V}^0$ that lie below the
worst equilibrium values; e.g., $\mathbf{V}^0=-\mathbf{D}^L$. Since we here consider full
failure costs,
$\mathbf{d}^A(\mathbf{V}^0) = \mathbf{0}$ and $\mathbf{b}(\mathbf{V}^0,\mathbf{p})
= (\mathbf{d}^A(\mathbf{V}^0)+\mathbf{p})\mathbbm{1}\{\mathbf{d}^A(\mathbf{V}^0)+\mathbf{p}<\mathbf{D}^L\}
= \mathbf{p}\mathbbm{1}\{\mathbf{p}<\mathbf{D}^L\} $.
That is, if we start from the pessimistic assumption that all banks are insolvent,
then they get no repayments from their debtors. Thus a bank is solvent in the following step of the
algorithm if and only if $p_i\geq D_i^L$, that is if and only if  it is unilaterally solvent.
If no bank is unilaterally solvent, the algorithm stops: assuming that all banks are insolvent is
self-fulfilling.

\paragraph{A Sufficient Condition for Iterative Solvency.}

One way to ensure having an iteratively strongly solvent set intersecting each directed cycle is to
have at least one unilaterally solvent bank on each cycle, but this is not always necessary, and
so the iterative solvency condition is important.
Nonetheless, an extended version of unilateral solvency is both necessary and sufficient whenever the banks that lie on multiple (simple) cycles
are critically balanced.

Let us say that a cycle is {\sl maximal} if there is no larger cycle containing all the banks in the original cycle.

\begin{corollary} \label{critically_balance}
If all banks that lie on {multiple} (simple) cycles are critically balanced, then all banks are solvent in the worst equilibrium \emph{if and only if} the network is weakly balanced and every (simple) cycle has at least one unilaterally solvent bank presuming that all debts owed into each maximal cycle are paid.
\end{corollary}

An implication of Corollary \ref{critically_balance} is that the iterative portion of the iterative solvency condition
 only matters when some of the banks that lie at the intersection of several cycles
have capital buffers, so that they only need some of their incoming debts to be paid before they
become solvent and can make their debt payments.
In that case,
those banks can be part of a repayment cascade in which payments in one cycle spread to another.
As an illustration, recall the network depicted in Figure \ref{fig:algorithm}. Bank 2 is the only
bank at the intersection of several (simple) cycles and is exactly balanced. In the worst equilibrium only
Bank 1 is solvent, as there is no unilaterally solvent bank on the right cycle.
Suppose instead that $p_2=1$. Bank 2 now has enough of a capital buffer
so that, even if it gets payments from only one cycle, it can make all of its payments.
All banks are then solvent in the worst equilibrium: though Bank 2 is not unilaterally solvent itself,
it has enough buffer so that the debt payment initiated by Bank 1 cascades and spreads to the right cycle.

\section{Minimum Bailouts Ensuring Solvency in any Given Equilibrium}\label{bailouts}

The results above characterize systemic solvency in the best and worst equilibria.
Next, we leverage these results to deduce the minimum bailouts
needed to return the whole network to solvency whenever there are some insolvencies.
These bailouts are the smallest transfers that avoid all failure costs and other inefficiencies
associated with a dysfunctional financial system,
and are optimal in this sense.\footnote{If anticipated, these bailouts
could distort banks' incentives and lead them to take on more risks ex ante.
Such moral hazard problems are well-studied, and so we do \emph{not} argue
that the regulator \emph{should} always ensure solvency ex post.
We simply acknowledge the fact that regulators often have to intervene to restore
solvency when faced with a crisis that has a potential for large cascades. This is thus an important problem,
and we analyze what is the most efficient way of doing so.}
Minimum bailouts
depend on the network structure:
just as losses can cascade and cycle through the network,
the same operates in reverse and well-placed bailouts
can have far-reaching consequences.

\subsection{The Minimum Bailout Problem}\label{minbail}

Consider a regulator who can inject capital $(t_1,\dots,t_n)\in \mathbb{R}^n_+$
into each bank in the network, increasing the value of bank $i$'s
balance sheet by $t_i$.
The timeline is as follows:
\begin{figure}[!h]
\begin{center}
\begin{tikzpicture}[scale=0.8]
\draw[->, thick] (0,0) to (11,0); \draw[thick] (0.5,0.2) to (0.5,-0.2);
\draw (0.5,-0.7) node {\small Returns $\textbf{p}$};
\draw (0.5,-1.2) node {\small are realized};
\draw[thick] (4,0.2) to (4,-0.2);
\draw (4,-0.7) node {\small Regulator};
\draw (4,-1.2) node {\small injects $\textbf{t}$};
\draw[thick] (10,0.2) to (10,-0.2);
\draw (10,-0.7) node {\small Values $ \textbf{V}(\textbf{p}+\textbf{t})$ and associated};
\draw (10,-1.2) node {\small  failure costs are realized };
  \end{tikzpicture}
  \end{center}
\end{figure}

First, returns on outside investments $\textbf{p}$ are realized.
Then, anticipating that some banks will be insolvent absent intervention at $\textbf{p}$,
the regulator can inject capital into the network.
Importantly, the capital is injected before banks officially start bankruptcy proceedings,
and hence before the associated failure costs are incurred.%

To ensure systemic solvency at minimum cost, the regulator solves
\begin{align*}
\label{opt}\tag{OPT}
&\min_{(t_1,\dots,t_n)\in \mathbb{R}^n_+} \quad \sum_i t_i\\[5pt]
&\text{s.t.}\quad \textbf{V}(\textbf{p}+\textbf{t})\geq \textbf{0}.
\end{align*}

Since there can be multiple equilibria for bank values,
we need to be precise about which equilibrium is selected in the constraint of (\ref{opt}).
For simplicity, we suppose that the transfers $\mathbf{t}$ do \emph{not} impact
the equilibrium selection, although they conceivably could.
Thus, to examine the minimum bailouts preventing defaults in the \emph{best}
equilibrium, we always select the \emph{best} equilibrium for bank values
$ \textbf{V}(\textbf{p}+\textbf{t})$ in (\ref{opt}). Similarly, if the goal is to prevent
defaults in the worst equilibrium, then optimal bailouts are the minimum transfers such that
all banks are solvent if we select the worst equilibrium for
$ \textbf{V}(\textbf{p}+\textbf{t})$.\footnote{More generally, fix any equilibrium
of $\textbf{V}(\textbf{p})$. Let $S$ the set of solvent banks in that equilibrium absent
intervention, and $N\setminus S$  the set of defaulting banks.
To keep the equilibrium selection
fixed, $\textbf{V}(\textbf{p}+\textbf{t})$ is computed by assuming
that banks in $S$ are
all solvent, and then finding the maximum set of defaults among the originally defaulting banks.}
Optimal
bailouts are
specific to an equilibrium. For instance, it can be that all banks are
solvent in the best equilibrium but some default in the worst. Optimal bailouts for
the best equilibrium are then null $\textbf{t}=\textbf{0}$ since all banks are already solvent,
but are strictly positive for the worst equilibrium.

We start by noting that Proposition \ref{solvency} yields a first characterization of
such minimum bailouts
for both the best and the worst equilibria, as well as other equilibria.

The best equilibrium is relatively easy to understand.
If the network is not weakly balanced then some banks must be defaulting, and each bank that is not
weakly balanced needs bailouts to be brought back to solvency.
It is thus necessary and sufficient for each bank $i$ to receive its net imbalance $t_i = [D_i^L-D_i^A-p_i]^+$ (by
Proposition \ref{solvency}). 

Since the minimum injections of capital that ensure systemic solvency in the best
equilibrium are fully characterized and relatively easy to calculate, for the remainder of the
paper we focus on the additional capital needed to ensure solvency in other equilibria.

Note that it is weakly optimal for the regulator to start by providing each bank
with its net imbalance $[D_i^L-D_i^A-p_i]^+$, as these payments are necessary for
solvency in all equilibria and  may trigger some repayment cascades. To analyze minimum bailouts for
an equilibrium other than the best, it is then without loss of generality to restrict
attention to the banks that remain insolvent in that equilibrium once net imbalances
have been injected, redefining their portfolio values to account for these transfers
and any debt payments they received from solvent banks.

Proposition \ref{solvency} then tells us that
the additional payments needed are the smallest ones that generate an iteratively solvent set of
banks that hits each defaulting cycle.

We first point out that even though (\ref{opt}) is written as a simultaneous choice
of payments, $\mathbf{t}\in \mathbb{R}^n_+$, it is equivalent to
specify an \emph{ordered} list of banks to bailout.
Indeed, any set of (simultaneous) payments $\mathbf{t}$ leads banks to become solvent in a particular order:
first some banks are made unilaterally solvent by the payments; then given their induced debt payments and the
bailout payments,
some other banks become solvent, and so forth. Reciprocally, any ordered list of bailouts
can be implemented via simultaneous payments by setting $t_i$ precisely equal to $i$'s
shortfall given the payments it gets from already solvent banks. Hence, just knowing which
banks get payments under the optimal bailout policy is not enough to characterize minimum
bailouts as the order in which banks are brought back to solvency matters.

To illustrate this, consider the network depicted in
Figure \ref{fig:complexity},
in which all banks default in the worst equilibrium.
Since Bank 2 lies on all cycles, bailing it out ensures systemic
solvency and costs $D_2^L-p_2=5$. This is however not the optimal policy, as the regulator
can significantly reduce the cost of bailing out Bank 2 if it first bails out Bank 3.
Indeed, bailing out Bank 3 costs $D_3^L-p_3=3$, and allows for the repayment of its debt to Bank 2.
Bailing out the latter then only costs $1$.
The optimal bailout policy is then $\mathbf{t}^* = (0,1,3)$,
which can be equivalently expressed as a sequence of banks that are bailed out  $\{3,2\}$.
\begin{figure}[!h]
\centering
\includegraphics[width=0.35\textwidth]{complexity_examples.tikz}
\caption{\small Let $p_1=0$, $p_2=5$, and $p_3=1$. 
}
\label{fig:complexity}
\end{figure}

As mentioned above, the order of bailouts matters, because the bailout cost of
remaining banks depends on which banks have already been bailed out.
Bailing out 3 and then 2 costs less than the reverse, and both lead to systemic solvency.

\subsection{The Computational Complexity of Finding Minimum Bailouts} \label{sec:complexity}

Proposition \ref{solvency}
implies that ensuring systemic solvency in equilibria other than the best
requires
that, beyond establishing weak balance, one also needs to inject enough capital so as
to ensure the existence of an iteratively strongly solvent set intersecting
all directed cycles on which banks are insolvent.
We examine this in what follows.

We begin by showing that this problem is
computationally complex, using concepts from the computer science literature.
Precisely, we prove that it is {\sl strongly NP-hard}. We provide formal definitions of such notions of complexity in Appendix \ref{sec:def}, but what matters is that every strongly NP-hard problem has instances for which finding exact or even approximate solutions
involves impractically many calculations (more than polynomially many in the size of the problem, based on known techniques).

\begin{proposition}\label{NPhard}
Finding whether there exists a bailout policy that ensures systemic solvency and costs no more than some budgeted amount is strongly NP-hard.
Thus, finding a minimum cost bailout policy (\ref{opt})
is also strongly NP-hard.\footnote{Checking whether
there exists a bailout policy that ensures systemic solvency and
costs no more than some budgeted amount
is an ``easier'' problem than finding a minimum cost policy since
knowing an optimal policy enables
one to answer the question of whether it can be done within some budget.
Thus, showing that the decision problem is strongly NP-hard
 establishes that (\ref{opt}) is as well,
 and working with decision problems is a standard technique.
}
\end{proposition}

The complexity comes from the fact that some of the capital that a bank needs to become solvent can come from the debts paid by others, and so it can be cheaper to first bail out a bank's debtors rather than bailing it out directly.    The number of possible combinations that could be optimal explodes factorially in the number of banks.  Moreover, given that there are many cycles in interbank networks there are no obvious starting points and so there are many situations where it is not obvious how to focus on just a small number of those combinations.

We prove Proposition \ref{NPhard} by showing that for some network structures,
finding whether one can make all banks solvent at no more than a certain cost enables one to solve the
``Largest Minimal Vertex Cover Problem,'' which is known to be strongly
NP-hard and hard to even approximate.\footnote{ \cite{boria2015max} show that,
unless $P=NP$, this problem cannot be approximated by a polynomial time algorithm
within ratio $n^{\varepsilon-0.5}$, for any $\varepsilon>0$.}
Given some undirected graph, a vertex cover is a set of vertices that contains at least one endpoint of all edges in the graph. A vertex cover is minimal if it is not the superset of another vertex cover. It turns out that, for some networks, the optimal bailout policy consists in bailing out banks belonging to a minimal vertex cover of maximum size.

To be more specific,
consider any undirected graph, and interpret an
edge between $i$ and $j$ in that network as bilateral claims of
1 that $i$ and $j$ have on each other (i.e., $D_{ij}=D_{ji}=1$).
So any edge $ij$ generates a cycle between
banks $i$ and $j$ in the financial network. Let $p_i=p\in(0,1)$ for all banks,
such that the network is \emph{critically} balanced: Banks' outside assets are
not enough to absorb the default of any one of their counterparties.
We know that, to ensure solvency of all, the regulator must bailout at least
one bank per cycle, and so the set of bailed out banks forms a vertex cover.
We furthermore show that this vertex cover has to be minimal (as otherwise the
regulator is bailing out banks that need not be bailed out) and that overall
bailout costs sum up to the number of edges net of the value of outside
assets of all bailed out banks. Hence an optimal bailout policy consists in
bailing out banks in a minimal vertex cover of \emph{maximal} size, as this allows
the regulator to leverage the outside assets of more banks. This means any Largest
Minimal Vertex Cover Problem can be translated into a Bailout Problem on an appropriately
constructed network, and so Bailout problems are at least as hard as
Largest Minimal Vertex Cover Problems.

We emphasize, however, that finding the optimal bailout policy is, in more general networks,
even more complicated
than the Largest Minimal Vertex Cover Problem.  As banks are bailed out, they repay their debts and
change the balance sheets of each other,
thus altering the remaining network of insolvent banks. 
This adds another layer of complexity, and means that the number of bailout policies that the
regulator would have to compare absent a better algorithm is of the order of $n!$,
which very quickly gets too large for any computer to handle.
With just 15 banks the problem already involves trillions of possible bailout strategies,
and with 20 banks more than $10^{18}$.
Furthermore, as the order of bailouts matters, (\ref{opt}) is not easily expressed as a linear
program, and standard dynamic programming
algorithms that have proved effective for NP-hard problems do not directly extend.%
\footnote{This makes the bailout problem
an interesting class of problems for further study in complexity.  This problem is strongly NP-complete as it is easy to check whether a given bailout policy costs no more than a certain amount. Yet it provides an interesting twist on
well-studied problems given that combinations of bailouts changes
the cost and value of other bailouts. One can expand the problem to have different values for every ordering, but then the
linear problem has factorially many inputs as a function of $n$.}

As discussed above, finding the minimum bailouts ensuring solvency in the best equilibrium is easy, so the added complexity for non-best equilibria comes from the fact that bailouts need to clear all the cycles. Hence the complexity of the problem does not scale with the number of banks in the network \emph{per se}, but with the number of cycles, which can be huge. However, if the network happens to contain few cycles (e.g., it has a core-periphery structure with a small core), then the minimum bailout problem remains tractable.  We formalize this intuition in Appendix \ref{sec:time} and provide an upper bound on the number of calculations needed to find the optimal policy that is exponential in the number of cycles only, but not in the number of banks.

\subsection{An Upper Bound on Bailout Costs}\label{sec:bound}

Even though finding the optimal bailout policy can be complex,
we can provide an upper bound on the total injection of capital that
is necessary to bring the network to solvency.  As discussed above, we assume that net
imbalances  $\sum_i[D_i^L-D_i^A-p_i]^+$ have already been injected, so this is a bound on the additional injections
needed to prevent self-fulfilling defaults.
We also provide a simple  algorithm that ensures systemic solvency, and leads to total
bailout costs that never exceed this bound. This is helpful since we know from
Proposition \ref{NPhard} that the minimum bailout problem is hard to approximate, and
that no known algorithm performs well.

The algorithm that we propose leverages the idea that bailing out a bank $i$ has an indirect
value: it brings $i$'s counterparties closer to solvency, or makes them solvent
altogether.\footnote{The indirect bailout value differs from the concept of threat index of
Demange \citeyearpar{demange2016}, which captures the impact of marginally increasing a
bank's repayments on total debt repayments, assuming the set of defaulting
banks remains the same. A key difference is that our notion of indirect bailout value is
not calculated on the margin, and the value of bailing out a bank
propagates further if we account for induced changes in the solvency status of other banks.}

Define the first-step indirect bailout value of a bank $i$ as
$\sum_j\min\{D_{ji}, (D^L_j-p_j-\sum_{k\text{ solvent}}D_{jk})^+\}$: it captures
by how much $i$'s solvency reduces the bailout cost of other banks in the network.
Note that this is a near-sighted measure of indirect bailout value as it
only accounts for a bank's \emph{direct} payments, but not for
the fact that $i$'s solvency can induce the solvency of one of its creditors $j$,
which then repays its debts, and so forth.
We can thus define a $k$th-step indirect bailout value of bank $i$ by all the cascades of
payments that are induced by the bailout of $i$ up to $k$ iterations of induced solvencies.
We provide formal definitions in the appendix, together with the proof of
Proposition \ref{bound_cost}.

When deciding whether to bailout $i$, the regulator must trade-off a bank's indirect bailout value with the cost
of its bailout, $(D_i^L-p_i-\sum_{j\text{ solvent}}D_{ij})^+$.
Consider the greedy algorithm that bails out the defaulting bank with highest
ratio of its $k$th-step indirect bailout value to its bailout cost
(for any $1\leq k\leq n-1$) until all are solvent, recomputing these
values after each step to account for all new
solvencies.\footnote{If there are any ties -- i.e., more than one bank with the highest indirect
bailout value to bailout cost ratio -- then break the ties arbitrarily.}
We show that this algorithm
leads to total bailout costs that never exceed the following bound.


\begin{proposition}\label{bound_cost}
For any $1\leq k\leq n-1$, bailing out banks in decreasing order of the ratio of their $k$th-step
indirect bailout value over bailout cost
until all are solvent
leads to a cost of at most
\[ \frac{1}{2}\sum_{i} (D_i^L-p_i)^+.\]
Thus, the optimal bailout cost
is no more than this, and
this bound is reached in some networks.
\end{proposition}

When following this algorithm, the regulator is guaranteed to inject at most
half of banks' total shortfall in order to ensure solvency of all banks.
This bound holds for all equilibria, 
and can be made tighter by only summing over the banks that default in the equilibrium of interest, instead of summing over all banks.
The bound is reached whenever the network is composed of $n/2$
disjoint cycles, no bank is unilaterally solvent, and banks are all equally costly to bailout -- so $(D_i^L-p_i)^+=(D_j^L-p_j)^+$
for all $i,j$.

This upper bound is not
straightforward. Consider, for instance, a clique of critically balanced banks,
such that all banks have claims on each other and none of them has enough capital
buffer to sustain the default of some of its counterparties.
Then, to ensure systemic solvency in the worst equilibrium, the regulator has to bailout \emph{all banks but one}.
What Proposition \ref{bound_cost} shows is that, even in such a network, total bailout costs
do not add up to more than half of banks' total shortfall: the optimal policy leverages
the fact that payments from banks that have already been bailed out reduce the cost of future bailouts.

Nonetheless, Proposition \ref{bound_cost} does
not imply that following the above algorithm always finds an optimal bailout policy.
For instance, in the example from Figure \ref{fig:complexity} it would begin
by bailing out Bank 2 rather than Bank 3 and would overpay.
In other instances it could overpay by
arbitrarily large amounts and so does not approximate the
optimal bailout policy well, as we show in Section \ref{sec:star} below.
The same is true of an algorithm that chooses banks in the order of the
indirect values that their bailouts generate.%
\footnote{Maximizing the indirect value divided by the
cost, or just the indirect value,
can lead the regulator to bail out the center bank in a star network if it has the highest debt liability/cost ratio,
while bailing out some peripheral banks can lead to (arbitrarily) lower total bailout costs.}

Similarly, algorithms that bail out banks in increasing order of their bailout
cost can also lead to total injections of capital arbitrarily larger than necessary.
For example, the network in Figure \ref{fig:greedy} illustrates how badly a greedy algorithm based on
minimizing costs can perform.
The example consists of a chain of cycles of length $n$. Note that banks $i\geq 2$ have enough
capital buffer to absorb the default of their smallest debtor, that is $i+1$. Hence if a bank $i$
is solvent, then this is enough to make its follower $i+1$ solvent as well, which is then enough
to make $i+2$ solvent, and this unravels until Bank $n$. The reverse is, however, not true:
Bank $i$ repaying its debts is not enough to make its predecessor $i-1$ solvent. Given this
asymmetry, the optimal policy is to bailout Bank 1: this guarantees solvency of all for a total
cost of $\overline{D}$. Now suppose the regulator uses a simple greedy algorithm, which consists in
bailing out banks in increasing order of their bailout cost. The algorithm bails out Bank $n$ first,
as this only requires injecting $D^L_n-p_n=\underline{D}<\overline{D} = D_i^L-p_i$ for all $i\neq n$.
The algorithm then bails out Bank $n-1$ at a cost of $D^L_{n-1}-p_{n-1}-D_{n-1n} = \overline{D} -\underline{D}$,
and then Bank $n-2$, etc. This leads the regulator to inject a total of
$\underline{D}+(n-1)(\overline{D} -\underline{D})$. Hence, the performance of the greedy algorithm
relative to the optimal policy can be arbitrarily bad for long enough chains.
\begin{figure}[!h]
\begin{center}
\begin{tikzpicture}[scale=1.2]
\definecolor{afblue}{rgb}{0.36, 0.54, 0.66};
\foreach \Point/\PointLabel in {(0,0)/1, (3,0)/2, (6,0)/3,  (12,0)/n}
\draw[ fill=afblue!40] \Point circle (0.4) node { $\PointLabel$};
\draw[ fill=afblue!40] (9,0) circle (0.4) node {{\footnotesize $n-1$}};
\draw[->, thick] (0.4,0.4) to [out=40,in=135] node[midway,fill=white]  {$\overline{D}$} (2.6,0.4);
\draw[<-, thick] (0.4,-0.4) to [out=-40,in=-135] node [ midway,fill=white]  {$\underline{D}$} (2.6,-0.4);
\draw[->, thick] (3.4,0.4) to [out=40,in=135] node [ midway,fill=white]  {$\overline{D}$} (5.6,0.4);
\draw[<-, thick] (3.4,-0.4) to [out=-40,in=-135] node [ midway,fill=white]  {$\underline{D}$} (5.6,-0.4);
\draw[->, thick] (6.4,0.4) to [out=40,in=135] node [ midway,fill=white]  {$\dots$} (8.6,0.4);
\draw[<-, thick] (6.4,-0.4) to [out=-40,in=-135] node [ midway,fill=white]  {$\dots$} (8.6,-0.4);
\draw[->, thick] (9.4,0.4) to [out=40,in=135] node [ midway,fill=white]  {$\overline{D}$} (11.6,0.4);
\draw[<-, thick] (9.4,-0.4) to [out=-40,in=-135] node [ midway,fill=white]  {$\underline{D}$} (11.6,-0.4);
  \end{tikzpicture}
  \end{center}
  \captionsetup{singlelinecheck=off}
  \caption[]{ Let $p_1=p_n=0$ and $p_i=\underline{D}$ for all $i\neq 1,n$.
  Suppose $\overline{D}>\underline{D}>0$.}
  \label{fig:greedy}
\end{figure}

As the example helps illustrate,
the complexity arises from the fact that cycles overlap, and that a repayment
cascade in one cycle can spread to another. More generally,
as banks are bailed out, they can either increase or decrease each other's indirect bailout values.
On the one hand, their bailouts and debt payments bring others closer to solvency and thus reduce how
much they still need, but then also make it easier to cause them to return to solvency
and induce subsequent payments.  So, a bank's bailout can affect
others' indirect bailout values in either direction.
This becomes especially complex when cycles overlap.

The optimal bailout problem is easier in networks in which
interferences between cycles are limited. This is for instance the case if cycles do not
overlap, and we fully characterize the optimal bailout policy in such networks in
Section \ref{sec:discycles}. This is also the case if all banks that lie on multiple cycles are
critically balanced -- i.e., have no capital buffer -- as then a repayment cascade cannot spread from one
cycle to another (Corollary \ref{critically_balance}).\footnote{See Online
Appendix \ref{sec:criticallybalanced} for an analysis of critically balanced networks.}

\section{Some Prominent Network Structures}

We showed that finding an optimal bailout policy can be both computationally hard, and hard to
approximate. In some financially relevant cases, however, the analysis simplifies and provides
insights regarding the structure of optimal bailouts. In this section, we first consider
networks in which cycles of banks do not overlap, and then examine networks that have a
core-periphery structure.  A main takeaway of our analysis is the importance
of indirect bailouts: to ensure the solvency of a large core bank, it is generally cheaper to
bailout small peripheral banks that owe money to it rather than to bailout the core bank directly, so as to leverage the outside assets of more banks. 

As in the previous section, we presume weak balance is satisfied and restrict attention
to the subnetwork of insolvent banks, accounting for all the debt payments that they
receive from solvent banks.

\subsection{Disjoint Cycles}\label{sec:discycles}

The first set of networks that we examine are those in which the cycles are disjoint, so that no
bank lies on more than one cycle.

There can still be banks that lie on no cycles, but for instance lie on directed paths coming out from
some bank on a cycle.   There
can also exist a directed path that goes from one cycle to another, as long as there is no directed
path that goes back
(which would violate each bank lying on at most one cycle).\footnote{Note that there cannot exist
banks with no incoming debts but with outgoing debts, since those
are already solvent under weak balance.}

Let  $c_1, \ldots, c_K$ be the simple cycles in the network,
where $c_k$ denotes the set of banks in the $k$-th cycle.
Order the cycles so that if $i\in c_k$ and $j\in c_{k'}$ for $k'>k$,
then there is no directed path from $j$ to $i$.\footnote{Given that $i$ and $j$ lie on different cycles,
there cannot be both a directed path from $i$ to $j$ and one from $j$ to $i$, as that would imply that
these banks lie on multiple cycles. Thus, there must exist such an
order, and in fact there can be multiple such orders.  Pick any one satisfying this condition,
as all such orders will lead to the same bailout costs.}

The optimal bailout policy is described as follows.
Pick the bank on $c_1$ that is the cheapest to bailout:
\[ \min_{i\in c_1} D_i^L-p_i.\]
This then ensures that all banks on that cycle are solvent, and can also lead to further solvencies as the banks on that cycle may owe debts to banks outside of that cycle.
Let $S_1$ denote the set of banks that are made solvent if all banks on $c_1$ are solvent (which is the same regardless of which bank on $c_1$ is bailed out).
If this includes banks on any other cycles, then that cycle will be fully solvent as well.
Consider the smallest $k$ such that $c_k\cap S_1=\emptyset$.   Accounting for all the payments that have come in from $S_1$, find the cheapest bank to bailout on $c_k$:
\[ \min_{i\in c_k} D_i^L-p_i-\sum_{j\in S_1} D_{ij}.\]
As before, let $S_2$ denote the set of banks made solvent by this bailout and its cascade.
Iteratively, after $h$ such steps, one finds the lowest indexed insolvent cycle $k_h$ and finds the cheapest bank on that cycle to bailout:
\[ \min_{i\in c_{k_h}} D_i^L-p_i-\sum_{j\in S_1\cup \cdots S_h} D_{ij}.\]
After at most $K$ steps the full network will be solvent.

\begin{proposition}\label{disjointcycles}
Suppose that each bank lies on at most one cycle.
Then, the optimal bailout policy consists of bailing out the bank closest to solvency on each simple cycle,
accounting for bailouts of previous cycles, as described above.
\end{proposition}

\subsection{Star Networks}\label{sec:star}

A star network is composed of one center bank, Bank $n$, that is exposed to all banks
in the periphery: $i\in\{1,\dots, n-1\}$. Each bank $i$ in the periphery is exposed only
to the center bank. A star network is then a simple example of a core-periphery network,
in which there is only one core bank. We first consider star networks as they provide
part of the intuition needed for the analysis of more general core-periphery networks.

The optimal bailout policy is more tractable in star networks than in some other
networks as only one bank, the center bank,
lies on multiple (simple) cycles. This reduces the complexity of the problem: finding the optimal bailout
policy is no longer \emph{strongly} NP-hard and becomes equivalent to the
Knapsack Problem.\footnote{A Knapsack
Problem consists of $n$ items that each have a weight and a value, and a knapsack that has some
limit on the total weight it can hold. The goal is to select which items to pack so as to
maximize the total value, while not exceeding the weight limit. Our problem is similar as,
intuitively, each peripheral bank has a value -- the amount of liquidity flow into the core
induced by its solvency,-- and a cost -- how much capital the regulator needs to inject to make it solvent.
The goal is then to minimize the total cost
of bailed out peripheral banks while inducing sufficient
liquidity flow into the core to ensure its solvency.} The Knapsack Problem is still an
NP-hard problem, but there exist several algorithms that approximate it arbitrarily well.
Hence our analysis suggests that these types of algorithms should be considered when
deciding which peripheral banks to bailout in core-periphery networks.

If the star network also satisfies some symmetry conditions,
then the optimal bailout policy can be fully characterized and needs not be approximated.
Let all peripheral banks be symmetrically exposed to the center bank:
they all have a debt claim of $D_{in}=D^{out}$ on the center bank,
and a liability $D_{ni} = D^{in}$ to it. So the total debt assets and liabilities of
the center bank equal  $D_n^A = (n-1)D^{in}$ and $D_n^L = (n-1)D^{out}$, respectively.
As before, let the network be weakly balanced,
and suppose none of the banks are unilaterally solvent.
If the center bank were unilaterally solvent, the whole network would always clear,
and regulatory intervention would be unnecessary.
Similarly, if some peripheral banks were unilaterally solvent,
then they could pay back their debt to the center bank for sure, and we
could redefine the network to account for these  payments.
See Figure \ref{fig:star} for an example of a star network.

\begin{figure}[!h]
\begin{center}
\begin{tikzpicture}[scale=0.8]
\definecolor{chestnut}{rgb}{0.8, 0.36, 0.36};
\definecolor{afblue}{rgb}{0.36, 0.54, 0.66};
\foreach \Point/\PointLabel in {(0,0)/2, (6,0)/3, (3,5.2)/1}
\draw[fill=afblue!40] \Point circle (0.45) node {$\PointLabel$};
\draw[fill=chestnut!40] (3,1.75) circle (0.45) node {$4$};
\draw[->, thick] (3.17,2.35) to node[near start,fill=white]{1}  (3.17,4.6); \draw[<-, thick] (2.83,2.35) to node[near end, fill=white]{2} (2.83,4.6);
\draw[->, thick] (0.45, 0.55) to node[near start, fill=white]{2} (2.4, 1.65); \draw[<-, thick] (0.6, 0.25) to node[near end,fill=white, below=-0.4em]{1} (2.55, 1.35);
\draw[<-, thick] (5.55, 0.55) to node[near end,fill=white]{1}   (3.6, 1.65); \draw[->, thick] (5.4, 0.25) to node[near start, fill=white, below=-0.4em]{2} (3.45, 1.35);
  \end{tikzpicture}
  \end{center}
  \captionsetup{singlelinecheck=off}
  \caption[]{\small A star network with three peripheral banks, $D^{in}=2$, and $D^{out}=1$.}\label{fig:star}
\end{figure}

Bailing out the center bank costs $(n-1)D^{out}-p_n$ and clears the whole system,
as the center bank lies on all cycles.
However, unless all peripheral banks have $p_i=0$, this is not optimal and can be much
costlier than optimal.

To understand the optimal bailout policy, let
\[
m^*=\frac{(n-1)D^{out}-p_n}{D^{in}}.
\]
Without loss of generality, let the peripheral banks be indexed in
decreasing order of $p_i$.

The optimal policy always involves first bailing out the $\lfloor m^* \rfloor$ peripheral banks
with the highest
portfolio values.
This leverages their $p_i$'s and gets  an amount of liquidity $\lfloor m^* \rfloor D^{in}$
into the center bank at a potentially
much lower cost than paying it directly to the center.
This then just leaves a comparison between the marginal shortfall of the center bank
\begin{equation*}
(n-1)D^{out} - p_n - \lfloor m^* \rfloor D^{in}
\end{equation*}
and the cost of bailing out the
$\lfloor m^* \rfloor + 1$-st peripheral bank
\[
D^{in}-p_{\lfloor m^* \rfloor + 1}.
\]
It is optimal to do the remaining bailout via the center bank instead
of the $\lfloor m^* \rfloor + 1$-st peripheral bank if and only if\footnote{Note that if
$m^*$ happens to be an integer, then the center bank is already solvent. }
\begin{equation*}
(n-1)D^{out} - p_n - \lfloor m^* \rfloor D^{in}
\leq
D^{in}-p_{\lfloor m^* \rfloor + 1}.
\end{equation*}

\begin{proposition}\label{star}
The optimal bailout policy is:
to bailout the first $\lfloor m^* \rfloor + 1$ peripheral banks if
\begin{equation*}
(n-1)D^{out} - p_n - \lfloor m^* \rfloor D^{in}
>
D^{in}-p_{\lfloor m^* \rfloor + 1},
\end{equation*}
and otherwise to bail out the first $\lfloor m^* \rfloor$ peripheral banks, and then
inject $(n-1)D^{out} - p_n - \lfloor m^* \rfloor D^{in} $ into the center bank.
\end{proposition}

The proposition implies that it is always best to start by bailing out peripheral banks,
so as to leverage their outside assets and reduce the overall cost of bailouts.
Then, once only one more peripheral bank's payment
is needed to return the center bank to solvency,
the regulator either injects the center bank's marginal shortfall or bails out one last peripheral bank,
depending on which  is cheaper.

An implication of Proposition \ref{star} is that mistakenly bailing out the center bank
when the optimal policy
is to target peripheral banks can be much costlier to the regulator than the reverse.
Indeed, this can lead the regulator to waste the value of the outside assets owned by peripheral banks.
In contrast, the only reason why bailing out peripheral banks may not be optimal is
if this leads the regulator to bailout  ``one too many'' of them due to the integer constraint,
which implies a waste of capital of at most $D^{in}$.

\begin{corollary}
Bailing out all peripheral banks leads the regulator to inject {at most}
an extra $D^{in}$ compared to the total amount required by the optimal bailout policy.

Bailing out the center bank leads the regulator to inject {at most}
an extra $\sum_{i=1}^{\lceil m^*\rceil}p_i $ compared to the total amount required by the optimal bailout policy.
\end{corollary}

Thus, in situations where the portfolio values among the needed peripheral banks
($\sum_{i=1}^{\lceil m^*\rceil}p_i $) is much larger than the amount any one of them owes to the center
bank ($D^{in}$), bailing out the periphery is a much safer policy in terms of wasted capital.

\subsection{Core-Periphery Networks and Cliques}\label{sec:core-periphery}

Many financial systems
are well-approximated by a core-periphery structure:
they can be decomposed
into a set of densely connected core banks and a
set of more sparsely connected peripheral banks.
In core-periphery networks, the above approach of beginning with peripheral
banks carries over, but then the remaining bailout problem of the core itself can be substantially
more complex.
We begin with cases in which there is some symmetry in the core, and in which intuitions can be drawn
 and optimal bailouts fully characterized. We then investigate the asymmetric case in the numerical simulations.

Let $N_C$ and $N_P$ be the set of core and peripheral banks, respectively,
with $N=N_C\cup N_P$ and $N_C\cap N_P=\emptyset$. The $n_C= \# N_C$
banks in the core form a  clique and are symmetrically exposed to each other:
$D_{ij} = D^{core}$ for all $i\neq j\in N_C$.
Each core bank $i\in N_C$ is also exposed to a different
subset of the peripheral banks $N_P^i\subseteq N_P$,
and each of these peripheral banks are only exposed to the core bank $i$.
Hence $N_P=\cup_{i\in N_C} N_P^i$
and $\cap_{i\in N_C} N_P^i=\emptyset$. Specifically, peripheral banks in $N_P^i$ have a
debt claim of $D^{out}$ on $i$, and a debt liability of  $D^{in}$ to $i$.
Suppose also that each core bank $i$ is exposed to the same number of peripheral banks  $n_P= \# N_P^i$ for all $i$.

The portfolio values of core banks are heterogeneous and denoted by $p_i$ and,
for simplicity, the smaller peripheral banks all have portfolios valued at $p_P$.

If the network is a clique, so that there is no peripheral banks and $n_P=0$,
then the optimal bailout policy is straightforward:
it simply consists in bailing out banks in decreasing order of their portfolio
value $p_i$ until they are all solvent.

The added complication from the periphery comes from the fact that peripheral banks change the bailout costs
of core banks.  However, note that we can combine the logic from
star networks, as in Proposition \ref{star}, with the logic from cliques.
In particular, consider the following bailout policy.

First, pick the core bank with the highest portfolio of outside assets $p_i$,
and let
\[
m^*_i=\frac{(n_C-1)D^{core}+n_PD^{out}-p_i}{D^{in}}.
\]
We can then follow the logic of Proposition \ref{star} to return this bank to solvency.
Start by  bailing out
$\lfloor m^*_i \rfloor$ of its peripheral banks, and then bail out one more of its
peripheral banks if
\[
(n_C-1)D^{core}+n_PD^{out}-p_i - \lfloor m^*_i \rfloor D^{in}
>
D^{in}-p_P,
\]
and inject the remaining
$(n_C-1)D^{core}+n_PD^{out}-p_i - \lfloor m^*_i \rfloor D^{in}$ directly
into bank $i$ otherwise.\footnote{If  $\lfloor m_i^* \rfloor \geq n_p$,
then first bail out all of $i$'s peripheral banks and then inject the remaining
amount into $i$ that is needed to make it solvent.}

Iteratively, if $x>0$ core banks have already been returned to solvency and the system is not yet at
full solvency,
then next consider the core bank with the highest $p_i$ among those still insolvent.
Define
\[
m^*_i(x)=\frac{(n_C-1-x)D^{core}+n_PD^{out}-p_i}{D^{in}}.
\]
Start by bailing out
$\lfloor m^*_i(x) \rfloor$ of its peripheral banks, and then bail out one more of its
peripheral banks if
\[
(n_C-1-x)D^{core}+n_PD^{out}-p_i - \lfloor m^*_i(x) \rfloor D^{in}
>
D^{in}-p_P,
\]
and otherwise inject the remaining
$(n_C-1-x)D^{core}+n_PD^{out}-p_i - \lfloor m^*_i(x) \rfloor D^{in}$ directly into bank $i$.\footnote{Again,
if $\lfloor m^*_i(x) \rfloor \geq n_p$, then bail out all of $i$'s peripheral banks, and then
inject the remaining amount into bank $i$ needed to make it solvent.}
Continue until all banks are solvent.

\begin{proposition}\label{coreperiphery}
The above policy is optimal.
\end{proposition}

Hence the intuition from Proposition \ref{star} extends to more general core-periphery networks:
it is always optimal to start by bailing out a core bank's peripheral counterparties instead of
targeting it directly.

For tractability, we imposed a symmetry assumption on the network: the only
heterogeneity came from the outside assets of core banks, and the difference between core and peripheral banks.
When banks are heterogenous within the core, then finding the optimal order
in which to bring core banks to solvency can be quite complex, and in fact can again
nest a vertex covering problem. Nevertheless,
as long as core banks are sufficiently large
compared to peripheral banks, it remains optimal to first target a core bank's peripheral banks instead of
bailing it out directly.  Hence one part of the optimal bailout policy is simple,
but the other part is not.\footnote{See Section \ref{asyexample} of the Online Appendix for an example of an asymmetric core-periphery network that highlights the tractability challenges of such networks.}  

\section{Comparing Bailout Policies: Simulations on Random Core-Periphery Networks}

To illustrate how various bailout policies perform in realistic settings, we focus on core-periphery networks, as many real-world financial networks have such a structure.
We show that results from the previous section, and in particular the cost-efficiency of bailing out some peripheral banks first, hold in a broader class of networks with various asymmetries that make analytical results challenging, but can be handled via simulations. We also derive new insights as to when ``na\"ive'' bailout policies perform particularly poorly compared to more sophisticated ones.

\paragraph{Simulated Random Networks.} As in Section \ref{sec:core-periphery}, the network is composed of $n_C$ core banks, which are all exposed to each other. Each core bank is also exposed to $n_P$ peripheral banks. 

Even though we fix the structure of interbank exposures---i.e., whether $D_{ji} >0$ for each $ij$---the strength of these exposures as well as the value of outside assets can vary across banks. Each simulated network is constructed by randomly drawing the liabilities ($D_{ij}$s) and outside assets ($p_i$s) from a uniform distribution between zero and one.
We vary how correlated these are across banks to explore how such asymmetries, which were precluded in Section  \ref{sec:core-periphery}, affect our results.

In each simulation, we construct the network ($\mathbf{D}, \mathbf{p}$) as follows:
\begin{enumerate}
\item We first construct one benchmark network ($\mathbf{D}^{\text{sym}}, \mathbf{p}^{\text{sym}}$) in which core banks are perfectly symmetric: they have the same portfolio of outside assets ($p^{\text{sym}}_i=p^{\text{sym}}_j\;\forall i, j\in N_C$), the same exposures to each other ($D^{\text{sym}}_{ij}=D^{\text{sym}}_{kl}\; \forall i, j, k, l \in N_C$) and isomorphic peripheries ($(D^{\text{sym}}_{ik},D^{\text{sym}}_{ki},p^{\text{sym}}_k)_{k\in N_P^i}=(D^{\text{sym}}_{jk},D^{\text{sym}}_{kj},p^{\text{sym}}_k)_{k\in N_P^j}\;\forall i, j\in N_C$).\footnote{All these are still drawn from a uniform distribution over $[0,1]$, e.g., we draw one number and set $p^{\text{sym}}_i$ equal to that number for all core banks $i$, etc.}
\item We then construct a second benchmark network ($\mathbf{D}^{\text{asym}}, \mathbf{p}^{\text{asym}}$) in which all the exposures ($D_{ij}$s) and outside assets ($p_i$s) are drawn independently across banks.
\item Finally, we construct our network of interest by taking a convex combination of the two benchmark networks: $(\mathbf{D}, \mathbf{p}) =\alpha(\mathbf{D}^{\text{asym}}, \mathbf{p}^{\text{asym}})+(1-\alpha)(\mathbf{D}^{\text{sym}}, \mathbf{p}^{\text{sym}})$.
\vspace{-0.2cm}
\end{enumerate}

The weight  $\alpha$ thus captures the level of asymmetries across core banks, as illustrated in Figure \ref{fig:simunetwork}.

\begin{figure}[!h]
 \centering
\begin{subfigure}[b]{0.25\textwidth}
      \centering
\begin{tikzpicture}[scale=0.8]
\definecolor{afblue}{rgb}{0.36, 0.54, 0.66};
\definecolor{chestnut}{rgb}{0.8, 0.36, 0.36};
  \path[use as bounding box] (0,3) rectangle (0.5,-3);
\foreach \x [count=\i] in {0,60,...,300}{
    \node[circle, draw, fill=chestnut!40,  inner sep = 0.6ex] (c\i) at (\x:1) {};
}
\foreach \x [count=\i] in {0,15,...,345}{
    \node[circle, draw, fill=afblue!40, inner sep = 0.6ex] (p\i) at (\x:2.5) {};
}
\foreach \x [count=\i] in {1,...,6}{
    \foreach \y [count=\j] in {1,...,6}{
        \ifnum\j>\i
            \draw[line width=0.6pt] (c\i) -- (c\j);
        \fi
            }
}
\foreach \x [count=\i] in {2,...,6}{
\pgfmathtruncatemacro{\start}{(\x-1)*4}
 \pgfmathtruncatemacro{\end}{(\x-1)*4+3}
   \foreach \y [count=\j] in {\start,...,\end}{
        \draw[line width=0.6pt] (c\x) -- (p\y);
    }
}
\draw (c1) -- (p1); \draw (c1) -- (p2); \draw (c1) -- (p3); \draw (c1) -- (p24);
\end{tikzpicture}
\end{subfigure}
\hspace{1cm}
\begin{subfigure}[b]{0.25\textwidth}
      \centering
\begin{tikzpicture}[scale=0.8]
\definecolor{afblue}{rgb}{0.36, 0.54, 0.66};
\definecolor{chestnut}{rgb}{0.8, 0.36, 0.36};
  \path[use as bounding box] (0,3) rectangle (0.5,-3);
\foreach \x [count=\i] in {0,60,...,300}{
	    \foreach \y [count=\j] in {0.7411, 0.7232, 0.6196, 0.5249, 0.7121, 0.4837}{
            \ifnum\j = \i
    \node[circle, draw, fill=chestnut!40,  inner sep = \y ex] (c\i) at (\x:1) {};
        \fi
    }
}
\foreach \x [count=\i] in {0,15,...,345}{
	    \foreach \y [count=\j] in {0.5165, 0.4751, 0.4675, 0.6860, 0.6807, 0.5561, 0.6521, 0.6194, 0.6714, 0.7367, 0.5339, 0.7372, 0.5797, 0.6152, 0.5956, 0.4704, 0.6752, 0.6160, 0.4951, 0.7452, 0.7247, 0.7309, 0.4689, 0.5137}{
            \ifnum\j = \i
              \node[circle, draw, fill=afblue!40, inner sep = \y ex] (p\i) at (\x:2.5) {};
        \fi
    }
}
\foreach \x [count=\i] in {1,...,6}{
    \foreach \y [count=\j] in {1,...,6}{
   	 \foreach \z [count=\k] in {0.6496, 0.5443, 0.5288, 0.6882, 0.7128, 0.7084, 0.7676, 0.8941, 0.6031, 0.7701, 0.4721, 0.7681, 0.4748, 0.8200, 0.4647, 0.5638, 0.7329, 0.6312, 0.6468, 0.8435, 0.6501}{
      	\ifnum\j>\i
	 \pgfmathtruncatemacro{\correct}{21-(7-\i)*(8-\i)/2+\j-\i}
      	\ifnum\k=\correct
            	\draw[line width=\z pt] (c\i) -- (c\j);
       	 \fi
       	 \fi
	 }
    }
}
\foreach \x [count=\i] in {2,...,6}{
\pgfmathtruncatemacro{\start}{(\x-1)*4}
 \pgfmathtruncatemacro{\end}{(\x-1)*4+3}
   \foreach \y [count=\j] in {\start,...,\end}{
      	 \foreach \z [count=\k] in {0.6263, 0.6883, 0.7162, 0.5095, 0.5506, 0.4377, 0.6320, 0.4471, 0.6659, 0.8011, 0.8315, 0.7434, 0.6968, 0.4771, 0.8876, 0.6828, 0.6265, 0.5766, 0.6983, 0.5814}{
	 \pgfmathtruncatemacro{\temp}{(\k+3}
      	\ifnum\temp=\y
      	  \draw[line width=\z pt] (c\x) -- (p\y);
       	 \fi
 	 }
   }
}
\draw[line width=0.6386pt] (c1) -- (p1); \draw[line width=0.4245pt] (c1) -- (p2); \draw[line width= 0.4408pt] (c1) -- (p3); \draw[line width=0.5994pt]  (c1) -- (p24);
\end{tikzpicture}
\end{subfigure}
\hspace{1cm}
\begin{subfigure}[b]{0.25\textwidth}
      \centering
\begin{tikzpicture}[scale=0.8]
\definecolor{afblue}{rgb}{0.36, 0.54, 0.66};
\definecolor{chestnut}{rgb}{0.8, 0.36, 0.36};
  \path[use as bounding box] (0,3) rectangle (0.5,-3);
\foreach \x [count=\i] in {0,60,...,300}{
	    \foreach \y [count=\j] in {0.8823, 0.8464, 0.6391, 0.4498, 0.8242, 0.3674}{
            \ifnum\j = \i
    \node[circle, draw, fill=chestnut!40,  inner sep = \y ex] (c\i) at (\x:1) {};
        \fi
    }
}
\foreach \x [count=\i] in {0,15,...,345}{
	    \foreach \y [count=\j] in {0.4330, 0.3502, 0.3350, 0.7720, 0.7615, 0.5123, 0.7042, 0.6387, 0.7428, 0.8733, 0.4678, 0.8744, 0.5594, 0.6305, 0.5911, 0.3409, 0.7504, 0.6321, 0.3903, 0.8905, 0.8494, 0.8618, 0.3379, 0.4275}{
            \ifnum\j = \i
              \node[circle, draw, fill=afblue!40, inner sep = \y ex] (p\i) at (\x:2.5) {};
        \fi
    }
}
\foreach \x [count=\i] in {1,...,6}{
    \foreach \y [count=\j] in {1,...,6}{
   	 \foreach \z [count=\k] in {0.6992, 0.4885, 0.4575, 0.7763, 0.8256, 0.8168, 0.9352, 1.1882, 0.6062, 0.9403, 0.3442, 0.9362, 0.3497, 1.0400, 0.3294, 0.5276, 0.8658, 0.6623, 0.6936, 1.0870, 0.7003}{
      	\ifnum\j>\i
	 \pgfmathtruncatemacro{\correct}{21-(7-\i)*(8-\i)/2+\j-\i}
      	\ifnum\k=\correct
            	\draw[line width=\z pt] (c\i) -- (c\j);
       	 \fi
       	 \fi
	 }
    }
}
\foreach \x [count=\i] in {2,...,6}{
\pgfmathtruncatemacro{\start}{(\x-1)*4}
 \pgfmathtruncatemacro{\end}{(\x-1)*4+3}
   \foreach \y [count=\j] in {\start,...,\end}{
      	 \foreach \z [count=\k] in {0.6526, 0.7767, 0.8323, 0.4190, 0.5013, 0.2754, 0.6640, 0.2942, 0.7319, 1.0023, 1.0631, 0.8868, 0.7937, 0.3542, 1.1751, 0.7656, 0.6530, 0.5532, 0.7966, 0.5627}{
	 \pgfmathtruncatemacro{\temp}{(\k+3}
      	\ifnum\temp=\y
      	  \draw[line width=\z pt] (c\x) -- (p\y);
       	 \fi
 	 }
   }
}
\draw[line width= 0.6773 pt] (c1) -- (p1); \draw[line width=0.2489 pt] (c1) -- (p2); \draw[line width= 0.2816 pt] (c1) -- (p3); \draw[line width=0.5987 pt]  (c1) -- (p24);
\end{tikzpicture}
\end{subfigure}

\caption{\small Sample networks with $n_C=6$ core banks, each exposed to $n_P=4$ peripheral banks, when asymmetries across core banks are $\alpha=0$ (left panel), $\alpha=0.5$ (middle), and $\alpha=1$ (right). To keep the graph readable, we only draw one link between any two counterparties, though there should be two (one pointing from $i$ to $j$ and one from $j$ to $i$). The size of a node $i$ captures the size of that bank's outside assets $p_i$ and the width of an edge $ij$ captures the associated exposure $D_{ij}$. }
\label{fig:simunetwork}
\end{figure}

For each constructed network, we first inject the net imbalances $t_i = [D_i^L-D_i^A-p_i]^+$ that guarantee solvency in the best equilibrium. Since we are mostly interested in the additional transfers needed to guarantee solvency in the worst equilibrium, we focus on those in what follows. Effectively, it is as if we had redefined banks' outside assets to be $\hat{p}_i=p_i+ [D_i^L-D_i^A-p_i]^+$.

\paragraph{Targeting the Core Directly vs. Starting with the Periphery.} We revisit and illustrate the key insight from Sections \ref{sec:star} and \ref{sec:core-periphery} that bailing out (some) peripheral banks first is often cost-efficient. For each simulated network, we compute the total bailout costs associated with three policies:
\smallbreak

\noindent\textsc{Core First}: We bail out core banks in decreasing order of their Indirect-Bailout-Value-

\hspace{2cm} over-Bailout-Cost ratio, recalculating ratios after each bailout.\footnote{We use the first-step indirect bailout value for simplicity.}
\smallbreak

\noindent\textsc{Periphery First}: We bring core banks back to solvency in the same order as under \textsc{Core}

\hspace{2cm} \textsc{First}, but, at each step, we check whether bailing out a subset of the core

\hspace{2cm} bank's periphery is cheaper than targeting that core bank directly.
\smallbreak

\noindent\textsc{Optimum}: We derive the optimal bailout policy.\footnote{Because finding the optimal bailout policy is computationally hard, we only consider networks of modest size in the simulations. }
\smallbreak

First, we examine how the  ``na\"ive policy" that targets core banks first performs relative to a more sophisticated policy that targets the periphery first, and how they both compare to the optimum. The optimal policy also targets the periphery first but following an optimal ordering over the core as discussed in Proposition \ref{coreperiphery}.  We compare these three policies as a function of the size of the periphery. For this analysis, we impose no symmetry on the core and set $\alpha=1$.

\begin{figure}[!h]
\centering
\includegraphics[width=0.45\textwidth, trim=50 10 50 15]{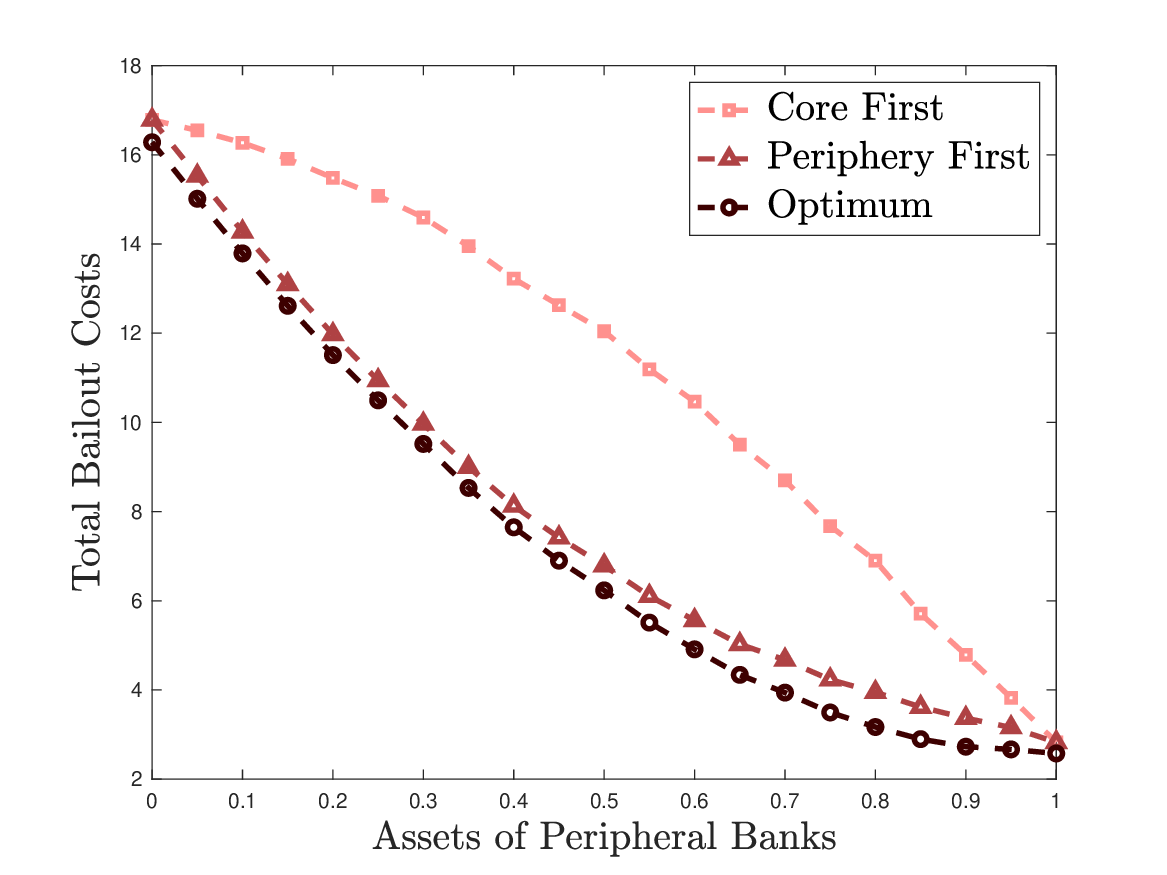}
\caption{ How bailout policies perform as a function of the size of peripheral banks (averaged over 1,000 simulations). For these simulations only, all peripheral banks have the same portfolio of outside assets $\hat{p}_i=p$, which is not drawn randomly but varied deterministically so as to plot total bailout costs as a function of $p$. Interbank exposures and core banks' outside assets are drawn i.i.d. from $U(0,1)$. ($n_C=6$, $n_P=4$, $\alpha=1$.) }
\label{fig:coreperipherysim}
\end{figure}

The answer depends nonmonotonically on the size of peripheral banks' asset holdings, as depicted in Figure \ref{fig:coreperipherysim}.
Indeed, \textsc{Periphery First} offers the largest relative improvement over \textsc{Core First} when peripheral banks have assets of intermediate size.
When peripheral banks have large outside assets, they are most likely solvent already, and there is little to no gain in considering them carefully in the bailout policy. When they have almost no outside assets, then they have nothing that the regulator can leverage to reduce bailout costs.
It is in the intermediate case that the regulator can gain a lot by targeting peripheral banks first. Doing so brings the regulator much closer to the optimal policy. The (relatively small) remaining gap with \textsc{Optimum} comes from the fact that \textsc{Periphery First} might bring core banks back to solvency in a suboptimal order.

\begin{figure}[!h]
\centering
\includegraphics[width=0.45\textwidth, trim=50 10 50 15]{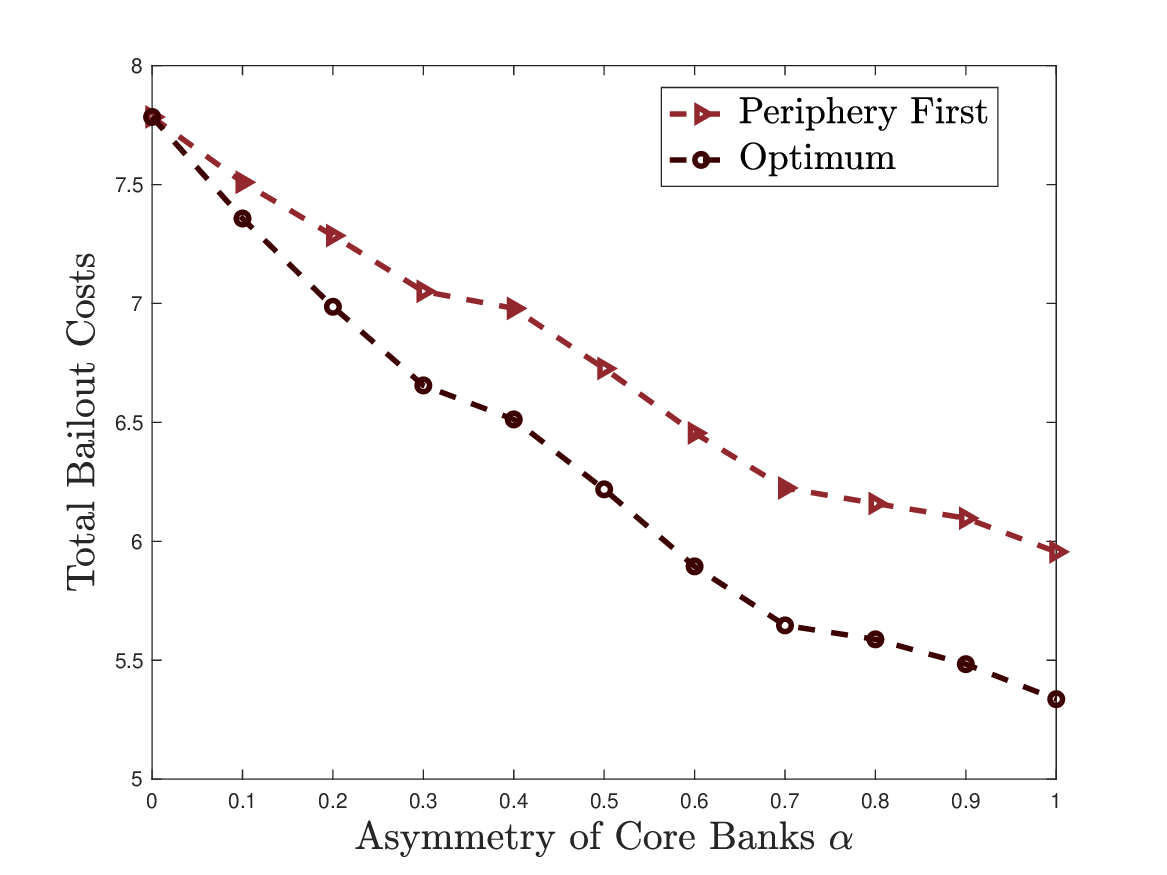}
\caption{
How the \textsc{Periphery First} bailout policy compares to the optimum as a function of the asymmetry among core banks.  The parameter $\alpha$ captures how asymmetric core banks are. At one extreme ($\alpha=1$), all interbank exposures and outside assets are drawn independently. At the other extreme ($\alpha = 0$), core banks are identical: they have the same exposures, assets, and peripheries. ($n_C=6$, $n_P=4$) }
\label{fig:simunetwork2}
\end{figure}

Next, we examine the remaining gap between \textsc{Periphery First} and the optimum bailout policy, as a function of how much asymmetry there is among core banks.
As discussed above, there is no systematic way of finding the optimal order in which to bail out the core. Core banks form a clique composed of many overlapping cycles, and absent some symmetry, the many cycles lead to a complex problem.  \textsc{Periphery First} is somewhat ``na\"ive'' in that it decides which core bank to bring back to solvency in a myopic way: it simply uses the greedy algorithm proposed in Section \ref{sec:bound} and brings core banks back to solvency in decreasing order of their indirect bailout value to bailout cost ratio.\footnote{As described above, \textsc{Periphery First} is still more sophisticated than \textsc{Core First} as, before each core bank's bailout, it checks whether bailing out a subset of that bank's periphery is cheaper than targeting that core bank directly.} The performance of such a policy very much depends on asymmetries across core banks, as depicted in Figure \ref{fig:simunetwork2}.

 If core banks are symmetric, there is no gain in ordering the bailouts of core banks in sophisticated ways, and the simple  \textsc{Periphery First} is optimal. However, when core banks are nontrivially asymmetric, the optimum bailout policy exploits these asymmetries and simpler bailout policies perform worse.

\section{Discussion}

\subsection{Recovery of Capital Injections}

Our analysis distinguishes between the minimum bailouts needed to ensure systemic
solvency in the best equilibrium, and those needed to avoid additional self-fulfilling
defaults in non-best equilibria. We showed that the former are easy to characterize as they
do not depend on the specific network structure and only require rebalancing each bank's portfolio,
whereas the latter depend on network cycles in complex ways.
Another feature that distinguishes bailing out the best equilibrium versus others
is that payments made to achieve weak balance cannot be
recovered by whomever intervenes in the bailout, at least in the short run.\footnote{This is in
the context of our static setting.  It could be that once returned to solvency, banks' future profits will
enable them to pay back current shortfalls. }
In contrast, additional payments made to ensure solvency in other
equilibria
can be recovered,
at least theoretically, once
they have cycled through the network.
These could thus be offered, for instance, as short-term loans.

Even though these  additional bailouts can, in theory,
be recovered,
doing so may be infeasible in practice for a variety of reasons.
Furthermore, even if the regulator does manage to recoup a large fraction of the bailout money,
injecting it requires a lot of capital ex ante, which can be costly in itself.
Hence finding the minimum bailouts that ensure systemic solvency in the worst equilibrium is
policy relevant, even though some of the funds used in these bailouts can eventually be
recuperated.\footnote{See \cite{lucas2019measuring} for a detailed look at the total bailout money
injected, and the amount that was not recovered, in the 2008 crisis.}
An extension of our analysis is to assess the risks of recovering funds from different banks, and
accounting for those differences in the cost of a bailout,  so that each
bank has an adjusted cost of funds injected into it.

\subsection{Bailouts with a Limited Budget}\label{budget}

We considered the problem of ensuring solvency of {all} banks at
{minimum} cost to the regulator. We showed that ensuring systemic solvency in the best
equilibrium is fairly simple, and hence focused on studying non-best equilibria.
In some settings, however, the regulator faces a related but distinct problem: it has
a fixed bailout budget $W$ to allocate across banks and wants to minimize
defaults and deadweight losses in some equilibrium of interest.
If the budget $W$ is low enough such that not all defaults
can be avoided, then
many of the results and intuitions that we derived for non-best equilibria
also apply to the best equilibrium. In particular, finding a bailout policy that minimizes
defaults in the best equilibrium is (strongly) NP-hard when there is a limited budget.
Indeed, because of the budget constraint, there is a trade-off between the indirect value associated
with a bank's bailout and its cost, that parallels the analysis above. Finding optimal bailouts is again
challenging since the
order of bailouts matters, and becomes similar to
the above problem of finding the minimum amounts
of payments that bring, in this case the most, banks back to solvency.

The intuition underlying Proposition \ref{star} and \ref{coreperiphery} also extends to the
best equilibrium. Indirect bailouts are generally the most cost-efficient: bailing out a bank's
creditors first instead of targeting it directly leads to lower total bailout costs for the
same set of induced solvencies.

\subsection{Ex-post Renegotiations and Mergers}

Ex-post interventions, such as renegotiations and mergers, can attenuate the risk of miscoordination and associated inefficiencies. If, instead of defaulting, a bank can renegotiate its payments down to what it can actually repay, or even merge with its counterparties, then failure costs can be avoided. For instance, if banks can merge costlessly, then coordination on the best equilibrium can be trivially achieved by having all banks merge to form a single entity such that no cycle remains. Such ex-post interventions are however unlikely to be frictionless. Contracts would have to be renegotiated multilaterally, to ensure each bank repays as much as possible given others' renegotiated contracts. In practice, efficient renegotiations would not only be computationally hard to achieve but also impeded by other frictions such as asymmetric information, as in Glode and Opp \citeyearpar{glode2021private}. 

Importantly, the extent to which efficient renegotiations and mergers occur depends on banks' incentives to intervene ex-post, which are shaped by the bailout policy chosen by the regulator. In particular, public bailouts might crowd out private bail-ins, as banks might not be willing to undertake costly negotiations if they anticipate the regulator to step in (Bernard, Capponi and Stiglitz \citeyearpar{bernard2017bail}).  The regulator might then benefit from committing to less extensive bailouts to reduce such crowding out.

\subsection{The Benefits of Netting, Compression, and CCPs}\label{compression}

Our analysis has made clear the role of cycles in generating
multiple equilibria for bank values.
This implies that there are potential gains from
clearing cycles in the network, as this can eliminate the possibility of a market
freeze and reduce this form of systemic fragility.
In practice, this resonates with liquidity-saving mechanisms as well as portfolio compression.
Liquidity-saving mechanisms are settlement systems that allow banks to condition their payment
on the receipt of another payment (Martin and McAndrews \citeyearpar{martin2008economic}).
When offsetting payments are in the queue, the system can clear them, thus preventing a potential
freeze in which none of the parties pays back the others.
Portfolio compression, which has become increasingly popular since the 2008 financial crisis,
allows banks to eliminate offsetting obligations with other
organizations, exploiting cycles in the financial
network.\footnote{In some cases, portfolio compression can however worsen the default cascade triggered by a shock,
if failure costs are low and banks heterogeneous enough (Schuldenzucker and Seuken \citeyearpar{schuldenzucker2019}).}

Nevertheless, portfolio compression is still limited in practice.
Compression services are currently only available for certain securities,
mostly derivatives, and do not encompass multilateral nettings of all types of obligations
between financial institutions. Even within markets for which compression services are available,
there are large opportunities for compression that are not currently exploited.
For instance D'Errico and Roukny \citeyearpar{derricor2019} estimate that up to three quarters of the
notional value in the CDS contracts traded by European institutions could be compressed.
This is all the more surprising as the complexity and entanglements of CDS contracts
contributed to the turmoil and freeze of interbank debt markets in 2007.\footnote{See
``Credit derivatives: The great untangling'', \textit{The Economist}, Nov 6th 2008.}

Several reasons help explain why financial markets are far from being fully compressed.
The network is complex, with many overlapping and long cycles, and so offsetting
obligations requires participation of many parties, with full opening of their books to
a common entity, and banks may fail to coordinate as needed.
Contracts also have staggered maturities, covenants, and priorities that further complicate netting
since they are not fully comparable, and thus have to each be priced and closed out.
Finally, banks may not want some positions to be fully
compressed or others to be created in the netting process, as they may
prefer maintaining relationships with specific
counterparties (D'Errico and Roukny \citeyearpar{derricor2019}).\footnote{Note also that some contracts
may provide incentives for counterparties to monitor the activities of another bank, and those
incentives can disappear with netting.}
Furthermore, banks may prefer to hold offsetting long-term debts with others,
as such cycles allow them to insure against un-contractable liquidity shocks
(Donaldson and Piacentino \citeyearpar{donaldson2017netting}).
Such preferences matter in practice: for example, 
some of the current providers
of compression services ask participants their tolerance to portfolio reconfiguration
beforehand and, naturally, this restricts the extent to which the network can be compressed.

In any case, our analysis provides a base to estimate the system-wide gains associated
with netting: for any distribution over returns to outside investments,
one can compute by how much compressing the network reduces overall expected bailout costs,
for any given equilibrium.\footnote{What probability distribution
a regulator wants to put across equilibria
is outside the scope of our model, but one might wish to
work with worst case as a benchmark.} As discussed above, there are benefits associated with
having cycles in the network that would have to be valued
and then would account for the other side of the calculation.

Finally, the use of Central Counterparty Clearing Houses (CCPs) can also help mitigate some of these issues, as the resulting star-like network eliminates many cycles.
More generally, considering
regulations that change the network is an important topic for further research, but requires
modeling the endogenous formation and benefits from the network, which is beyond the scope of our analysis (e.g., see \cite{erolv2018,erol2019}).
Regardless of the precise policy that one undertakes, developing and maintaining a more complete
picture of the network, and the portfolios of banks together with those of their counterparties,
is a necessary first step to improving crisis management.

\smallskip

\begin{spacing}{1.0}
\bibliographystyle{ecta}
\bibliography{financeNetworks}



\begin{appendices}

\section{Proofs} \label{appendixA}

Before beginning the proofs, we note that using (\ref{eq-bookvalue-bankruptcy-i})
we can rewrite the expressions for the debt value (\ref{debtvalue})
in a single equation:
\begin{equation}\label{debtvalue2}
d_{ij}(\mathbf{V}_j) = \frac{D_{ij}}{D_j^L} \min\left(D_j^L,\max\left\{V_j+D_j^L,0\right\}\right),
\end{equation}
which accounts for both possible regimes:  either $j$ is solvent and creditors
share the promised payment $D_j^L$, or it is not and they share the book value of $j$'s
assets net of failure costs $V_j+D_j^L$, if positive. (\ref{debtvalue2}) makes it more
transparent that $i$'s debt assets can be written as a function of only $\mathbf{V}_{-i}$, so that
we can write $d_i^A(\mathbf{V}_{-i})$ and then rewrite (\ref{eq-bookvalue-bankruptcy-i})
as
\begin{equation}
\label{eq-bookvalue-bankruptcy-i2}
V_i=p_i +   d_i^A(\mathbf{V}_{-i})-D^L_i  - b_i(\mathbf{V},\mathbf{p}).
\end{equation}

\noindent{\bf Proof of Proposition \ref{uniqueV}:}
(i) Let $\overline{V}$ and $\underline{V}$ be the best and worst equilibrium values of banks, respectively.
Since there is no dependency cycle, it has to be that a subset of banks, denoted $X_0$,
only
derive value from their outside investments: $D_i^A=0$ for all $i\in X_0$.
Thus, by the assumption on failure costs,
$\beta_i\left(V_i,\mathbf{p} \right)$  only depends on $V_i$ for $i\in X_0$.
Therefore, the
 values of these banks are determined solely by their investments, can be written as:
\[V_i = p_i- D_i^L -  \beta_i\left(V_i,\mathbf{p} \right)\mathbbm{1}_{\{p_i < D_i^L\}}\quad \forall i\in X_0.\]
Since $ \beta_i\left(\cdot,\mathbf{p} \right)$ is a contraction,
there exists a unique solution to this equation by the Contraction Mapping Theorem.
Therefore, $\overline{V}_i=\underline{V}_i$ for all $i\in X_0$.
Next, let $X_1$ be the set of banks whose debt assets only involve banks in $X_0$.
Thus, their
value is independent of that of banks outside $X_0$, and so by (\ref{debtvalue2}) and (\ref{eq-bookvalue-bankruptcy-i2})
we can write
\[
V_i = p_i+\sum_{j\in X_0}d_{ij}(V_j) - D_i^L-  \beta_i\left((V_j)_{j\in X_0\cup\{i\}},\mathbf{p} \right)
\mathbbm{1}_{ \{ p_i +\sum_{j\in X_0} d_{ij}(V_j) < D_i^L \}
}
\quad \forall i\in X_1.
\]
Since $\overline{V}_i=\underline{V}_i$ for all $i\in X_0$ and $ \beta_i\left(\cdot,\mathbf{p} \right)$ is a contraction as a function of $V_i$, it follows that
the best and worst equilibrium values of banks in $X_1$ are also the same: $\overline{V}_i=\underline{V}_i$ for all $i\in X_1$.
Iteratively defining $X_k$ to be the set of banks that have debt claims on banks in $\cup_{j< k} X_{j}$ only, the same argument applies by induction.
For some integer $K\leq n$, $\cup_{k=0}^K X_k = N$, and thus it follows that $\overline{V}_i=\underline{V}_i$ for all banks.
\medskip

(ii) Next, we prove that if there is a dependency cycle in the network, then there exists
$\mathbf{p}$ and failure costs such that $\overline{V}\neq\underline{V}$.
In particular,
we first show that under failure costs from (\ref{canon}) with full failure costs (that is, when
a bank loses all of its assets upon default, $a=1$, $b=0$), then there exists returns $\mathbf{p}$ such that this is true.
We then show that this also holds for $a$ below 1 (and above some threshold).

Let $c$ be the set of all banks belonging to a
dependency cycle. All other banks either (i) owe debts that have value (directly or indirectly) flowing into a dependency cycle; (ii) get value from debts coming out of a dependency cycle;  or (iii) none of the above. Equilibrium values of banks in $c$ is independent from that of banks in categories (ii) and (iii),
because there are no directed paths of debts from such banks that are eventually owed to any bank in $c$.
The values of banks in (i) can impact those of banks in $c$,
but we set their portfolio values ($p_i$s) to zero and thus banks in $c$ get no value from them.\footnote{Note that none of these banks are part of a cycle, and so by the argument above their values are uniquely tied down, and since they have no debts coming from any banks in (ii) or (iii), it is direct that they have no assets with which to pay any of their debts.}
Thus, equilibrium values of banks in $c$ depend only on portfolios and values of banks in $c$, and each one of them has a debt  liability to at least one other and a debt asset from
at least one other.

Therefore, we redefine $D_i^A\equiv \sum_{j\in c} D_{ij}$ for all $i\in c$.
We set portfolio values $(p_i)_{i\in  c}$ to be as low as possible, while ensuring all banks in $c$ remain solvent in the best equilibrium.  In particular, when all banks in $c$ are solvent
their equilibrium values are
\[V_i = p_i+D^A_i-D_i^L,\]
and therefore the smallest $(p_i)_{i\in  c}$ that ensure they all remain solvent in the best equilibrium are given by
\[
p_i = [D^L_i-D_i^A]^+.
\]
Thus, by using these portfolio values all banks in $c$ are solvent in the best equilibrium: $\overline{V}_i\geq 0$ for all $i\in c$.

Next, consider what happens in the worst equilibrium with these portfolio values, under full failure costs --
that is under (\ref{canon}) with $a=1$, $b=0$. Recall that all banks in $c$ must have
some debt liability and some debt claim -- i.e., $D_i^A>0$, and $D_i^L>0$ for all $i\in c$.
Let us suppose that they all default so that $d_i^A(\mathbf{V}_{-i})=0$ for all $i$.
Then given that $p_i = [D^L_i-D_i^A]^+$
\[
V_i = p_i  - D_i^L =  [D^L_i-D_i^A]^+  - D_i^L  =-\min\{D_i^A,D_i^L\}<0\quad \forall i\in c,
\]
and assuming all banks in $c$ default is self-fulfilling.
Then given these failure costs $b_i(\mathbf{V},\mathbf{p})= p_i$, and so $d_i^A(\mathbf{V}_{-i})=0$ is self-fulfilling,
and the best and worst equilibrium values differ.

To complete the proof, we note that the same is true for some $a<1$, with $a$ sufficiently close to 1.  Presuming all banks default:
\[
V_i = (p_i + d_i^A(\mathbf{V}_{-i}))(1-a)  - D_i^L ,
\]
with $d_i^A(\mathbf{V}_{-i})$ continuous in $\mathbf{V}_{-i}$, as any discontinuity arises only when new defaults occur.
This has a unique fixed point when $a=1$ of $V_i =  - D_i^L$, and given that the righthand-side above is continuous in $\mathbf{V}_{-i}$ and $a$, then the fixed points converge to the limit
fixed point as $a\rightarrow 1$. So under the presumption that all banks default there is a solution for $a$ close enough to 1 with all $V_i<0$,
justifying that all banks default in the worst equilibrium and have values below 0,
differing from the best equilibrium.

(iii) First, we know that the set of equilibria forms a complete lattice,
and so banks that default in the best equilibrium must be defaulting in all other equilibria.
Second, the set of banks that default in a non-best equilibrium but not in the best
equilibrium must contain all banks from at least one cycle. By contradiction, suppose not; i.e., there
is no cycle of banks among these additional defaults. Then some of these banks must have claims
only on banks whose solvency status is the same in the non-best equilibrium and in the best equilibrium,
that is claims on banks who are either solvent in both equilibria, or insolvent in both equilibria.
Hence the value of these banks' assets is the same in these two equilibria, and they cannot be
defaulting in one but not in the other.

We are left to prove that any additional defaults must lie on outpointing paths from the original
or the newly defaulting banks. Note that a bank that does not lie on such outgoing path must be
getting the same value from its debtors in both equilibria. Hence it cannot be defaulting in the
non-best equilibrium while remaining solvent in the best equilibrium. \eproof

\bigskip

\noindent{\bf Proof of Proposition \ref{solvency}:}
First, note that weak balance is a necessary condition for a bank to be solvent in any equilibrium.   That is,
$ p_i+D^A_i<D^L_i$ for some $i$, then it must default in every equilibrium,
since it then defaults even if it gets all its incoming debts paid, and thus regardless of the solvencies of other banks.

Next, note that if $p_i+D^A_i-D^L_i\geq 0$ for all $i$ -- i.e., the network is weakly balanced -- then it is an equilibrium for all banks to be solvent, and so the best equilibrium
has all banks solvent.   Thus, weak balance is necessary and sufficient for solvency of all banks in the best equilibrium.
\smallskip

The characterization of solvency in the worst equilibrium requires more work.
First, weak
balance of the network is also a necessary condition for full solvency in the worst equilibrium, as argued above.
Weak balance is, however, no longer sufficient.
Thus, for the following, we suppose that the network is weakly balanced, and show that all banks are solvent in the worst equilibrium if and only if there is an iteratively strongly solvent set intersecting each directed cycle.

We first show that having an iteratively strongly solvent set intersecting each directed cycle (in addition to
weak balance) implies all banks are solvent.

Suppose there is an iteratively strongly
solvent set that intersects each directed cycle, and call it $B$. By the definition of an iteratively strongly
solvent set, all banks in $B$ must be
solvent in the worst equilibrium, which means that there is at least one solvent bank on each cycle.
We prove that this implies all banks in the network are solvent by induction on the number of cycles in the network $K$.

If there are no cycles, then the best and worst equilibrium coincide (see the proof of Proposition \ref{uniqueV}),
and then by the arguments above weak balance alone guarantees solvency of all banks (in fact, then it is easy to check that $B=N$ along the iterative arguments discussed above).
If there is exactly one cycle, then consider some bank $i$ on that cycle that is in $B$.
This bank $i$ is solvent in the worst, and thus in every, equilibrium given that all of $B$ is solvent.
So, consider the following modified network.  For each $j$ reset $p_j$ to be $p_j+D_{ji}$, so presume that all debts from $i$ are paid.  Then reset $D_{ji}$ to be $0$.
Note that we have not changed the structure of any equilibrium since $i$ would have been solvent in every equilibrium, and also that weak balance still holds.
The new network has no cycles, and thus all banks are solvent in all equilibria.
More generally, consider a network with $n$ cycles and some bank $i\in B$ that lies on some cycle.
Via the same argument, we end up with a modified network that has the same equilibria and fewer cycles,
and so by induction all banks are solvent in all equilibria.

\smallskip
Finally, we argue that if there does not exist an iteratively strongly solvent set that intersects every cycle,
then some banks default in the worst equilibrium. First note that the
union of iteratively strongly solvent sets is also an iteratively strongly
solvent set: hence there exists a maximum one which, by assumption,
does not intersect every cycle and hence does not include all banks.
One can then check that the maximum iteratively strongly solvent set is
actually the set of solvent banks in the worst equilibrium: presuming those banks are all solvent and none of the others
are solvent is self-fulfilling, and so $N\setminus B$ are not solvent in the worst equilibrium.\eproof

\bigskip

\noindent{\bf Proof of Corollary \ref{critically_balance}:}  We first show necessity by contrapositive. Suppose there is a simple cycle $c_k$ that has no unilaterally solvent bank on it, after all the debts have been paid into the maximal cycle in which it is embedded. The defaults of all banks on $c_k$ is self-fulfilling, even if we assume that all other banks in the network are solvent. Indeed, banks that \emph{only} lie on cycle $c_k$ only have one debt claim, on their predecessor in the cycle. Since they are not unilaterally solvent by assumption -- that is $D_i^L>p_i$ -- they cannot be solvent unless they get their debt back from their predecessor. Furthermore, all banks that lie on several cycles are critically portfolio balanced, so they cannot be solvent either unless they get all their debt payments from within the cycle. Hence assuming all banks are $c_k$ default is self-fulfilling, as each one of them defaults if its predecessor in the cycle defaults. Banks on $c_k$ must hence default in the worst equilibrium if there is no unilaterally solvent bank on the cycle given payments in from outside of the maximal cycle.

We next prove sufficiency. Note that any network can be partitioned into a hierarchy of banks:
\begin{itemize}
\item N0: banks that are not in any cycle but lie on a sequence of debts flowing into some (maximal) cycle.
\item N1: banks that are not in any cycle and but lie on a sequence of debts flowing out from some (maximal) cycle.
\item N2: banks that are not in any cycle and and not in N0 or N1.
\item C1: banks on maximal cycles that have no inflows from any other maximal cycle.
\item Ck: iteratively in k, banks on maximal cycles not already classified that have inflows from maximal cycles in level Ck-1 or above, but not any other cycles.
\end{itemize}

Let us say that a bank in level Ck is $k-1$-unilaterally solvent if it is solvent presuming that all banks in N0 and C1 through Ck-1 make their debt payments, but no other banks.

Banks that are in N0 and N2 are solvent under weak balance.
This implies that all debts into banks in C1 are paid from banks outside of those maximal cycles.
Thus, there exists an iteratively strongly solvent set that intersects all simple cycles involving banks in C1.
Iteratively, this applies to banks in Ck, and then in a final step to banks in N1.\eproof


\bigskip

\noindent{\bf Proof of Proposition \ref{NPhard}:} We show that finding the
size of the largest minimal vertex cover, which is known to be a strongly NP-hard
problem,\footnote{This is slightly different from the usual
minimum (rather than largest minimal) cover problem,
but is still strongly NP-hard.  See \cite{boria2015max}.} can be reduced to the decision problem
associated with (\ref{opt}) for some specific network structures.

The largest minimal vertex cover problem is stated as follows:
Given a graph, a vertex cover is a set of vertices that
includes at least one endpoint of all edges in the graph.
A vertex cover is minimal if it is not a superset of another
vertex cover.  The problem is to find the maximum size of a minimal vertex cover.

We show that if we can solve the decision version of the bailout problem for
some specific network structures, then we can solve any instance of the
decision version of the largest minimal
vertex cover problem.

For any undirected graph, build an interbank network as follows:
all vertices are banks, and all edges represent bilateral claims of size 1; i.e.,
if $ij$ is an edge then $D_{ij}=D_{ji}=1$. Hence any edge $ij\in E$ in the network generates a
cycle of claims between Banks $i$ and $j$. Finally suppose $p_i=0.5$ for all $i$.

Note that the network is critically balanced: each bank's total liabilities equal its total
debt assets $D_i^A=D_i^L$, and their outside assets are not enough to compensate one missing
payment since $p_i<1$.  Hence all banks default in the worst equilibrium.

We first argue that an optimal bailout policy has to bailout all banks in a minimal vertex
cover. One bank on each cycle must be bailed out to ensure systemic solvency, so the set of
bailed out banks must form a vertex cover. If it does not, then it means one edge; i.e.,
one cycle, is not cleared by the policy as none of the banks involved in the cycle is bailed out.
For the bailout policy to be optimal, the vertex cover must be minimal. If it isn't, then there
exists a bank that is bailed out and whose counterparties are all bailed out as well.
Bailing out this bank is not necessary since it is made solvent by receiving payments
from all of its counterparties.

We next show that bailing out a minimal vertex cover of cardinality $m$ costs $|E|-0.5m$, where
$|E|$ is the number of edges,
and hence that a bailout policy must bailout a minimal vertex cover of maximum size to be optimal.
The bailout policy cannot cost less than $|E|-0.5m$, since all cycles must be cleared, which requires
injecting at least 1 into all $|E|$ cycles, and that the total value of banks' assets that the
regulator leverages is $0.5m$. Furthermore, the bailout policy cannot cost more than $|E|-0.5m$.
If it did, the regulator would be injecting more than 1 in some cycle, which cannot be optimal as
injecting 1 in all cycles ensures systemic solvency. Let $W$ be the minimum total bailout costs.
Hence there exists a minimal vertex cover of size $m$ if and only if $W\leq |E|-0.5m$:
if we can solve the bailout problem, then we can solve any
instance of the largest minimal vertex cover problem.\eproof

\bigskip

\noindent{\bf Proof of Proposition \ref{bound_cost}:}
We prove -- by induction on the number of banks $n$ -- that the total injection of
capital needed to ensure systemic solvency is no greater than
$0.5\sum_i (D_i^L-p_i)^+$.
We provide the proof for the worst equilibrium, which then ensures that this works for any equilibrium.
Also, we provide the proof for the first-step indirect value over cost ratio, and then discuss
how it extends to the $k$th-step indirect value over cost ratio, for $k\geq 2$.

Given the set of banks $S$ that are already solvent, let $x_{ij} \equiv \min\{D_i^L-p_i-\sum_{k\in S}D_{ik}, D_{ij}\}$ be the liquidity flow that $j$'s solvency induces in bank $i$. Let $x_j\equiv \sum_i x_{ij}$ be the overall indirect bailout value of bank $j$, that is the overall liquidity flow of interest that $j$'s solvency induces in the network. Finally, let $c_i \equiv (D_i^L-p_i-\sum_{j\in S}D_{ij})^+$ be the cost of bailing out bank $i$. Consider the algorithm that bails out the defaulting bank with highest $x_i/c_i$, recomputing these values after each step to account for all new solvencies.

It is simple to show that the claim holds for $n=2$, as there is then a single network configuration to
consider in which each bank has a claim on the other.\footnote{If not, then there is no
cycle in the network and all banks are already solvent as the network of interest is
assumed to be weakly balanced.}  Bailing out any of the two banks is enough to clear the cycle,
and so $x_1 = D_2^L-p_2$ and $x_2  = D_1^L-p_1$.
Then the algorithm bails out the bank closest to solvency,
which costs $\min\{(D_1^L-p_1)^+,(D_2^L-p_2)^+\}\leq 0.5[(D_1^L-p_1)^++(D_2^L-p_2)^+]$.

Now suppose the induction hypothesis holds for a network of $n$ banks, and consider a network with
$n+1$ banks.  Let $i^*=\arg \min_i x_i/c_i$ be the first bank picked by the algorithm, such that
$i^*$ is the bank with highest indirect bailout to bailout cost ratio. We give a bound on total
bailout costs if the regulator starts by bailing out $i^*$.

The regulator first bails out $i^*$ at a cost of $D^L_{i^*}-p_i^*=c_{i^*}$.
We can then restrict attention to the remaining network of $n$ banks, but we need to account for the payments they got from $i^*$. In aggregate, the remaining banks get a payment of $\sum_j\min\{D^L_{j}-p_{j}, D_{ji^*}\}=x_{i^*}$ from $i^*$. In that remaining subnetwork of $n$ banks, the induction hypothesis tells us that the optimal bailout policy costs at most $0.5\sum_{i\neq i^*} (D_i^L-p_i)^+ - x_i^*$. In total, accounting for the cost of $i^*$'s bailout, ensuring solvency then costs at most
\begin{align*}
0.5\sum_{i\neq i^*} (D_i^L-p_i)^+ -x_{i^*}+c_{i^*}.
\end{align*}
We then have to argue that $x_{i^*}\geq c_{i^*}$. By contradiction, suppose that $x_{i^*}/c_{i^*}<1$. By definition of $i^*$, this means $x_i/c_i<1$ for all $i$. However, we know that, since the network is weakly balanced, each banks getting its payments back in full has to be solvent. So
\[\sum_i x_{ij} = \sum_i \min\{D_j^L-p_j, D_{ji}\}\geq \min\{D_j^L-p_j, D_j^A\}\geq c_j.\]
Hence $\sum_j\sum_i x_{ij}\geq \sum_j c_j$, or equivalently $\sum_j x_j\geq \sum_j c_j$. There must then be at least one bank with $x_j\geq c_j$, and so $x_{i^*}\geq c_{i^*}$. The overall cost is then at most
 \begin{align*}
0.5\sum_{i\neq i^*} (D_i^L-p_i)^+ \leq 0.5\sum_{i} (D_i^L-p_i)^+,
\end{align*}
and the claim is true.

\paragraph{Extension to any $k$th-step indirect bailout value.}
As mentioned in the body of the paper, indirect bailout
values can be defined at various levels to accounts for cascades of indirect payments induced by a bank's solvency,
not only in terms of its payments, but further levels of solvencies and subsequent payments that its payments
induce. Let $I_i^0\equiv \{i\}$ and,
taking as given some set of banks $S$ that are already solvent, $i$'s bailout induces the solvency of
banks in
$$I_i^1\equiv \{j \notin S\cup I_i^0 | D_j^L-p_j\leq  \sum_{\ell\in S\cup I_i^0}D_{j\ell}\}.$$
These banks paying
back their debts leads to the following additional ``indirect'' solvencies
$$I_i^2\equiv \{j \notin S\cup I_i^0\cup I_i^1| D_j^L-p_j\leq \sum_{\ell\in S\cup I_i^0\cup I_i^1}D_{j\ell}\}.$$
Define recursively the set of banks made indirectly solvent at step $k$ as
$$I_i^k\equiv \{j \notin S\cup I_i^0\cup\dots\cup I_i^{k-1}| D_j^L-p_j\leq
\sum_{\ell\in S\cup I_i^0\cup\dots\cup I_i^{k-1}}D_{j\ell}\}.$$
This process must terminate at some step $k\leq n-1$, given the finite number of banks.
A bank $i$'s {\sl $K$-th step indirect bailout value}, for $ 2\leq K\leq n-1$, is then
\begin{align*}
\sum_{j\notin S\cup I_i^0}  \min\{ D_{ji}, D_j^L-p_j-\sum_{l\in S}D_{jl}\}+\sum_{k=1}^{K-1}
\sum_{h\in I_i^k} \sum_{j\notin S\cup I_i^0\cup\dots\cup I_i^{k}}
\min\{ D_{jh}, D_j^L-p_j-\sum_{m\in S\cup I_i^0\cup\dots\cup I_i^{k-1}}D_{jm}\}.
\end{align*}
The proof of Proposition \ref{bound_cost} works similarly for any $k$th-step indirect bailout value.
The case with $n=2$ is unchanged as these notions of indirect bailout value are all equivalent with
only two banks. Note that a bank's $k$th-step indirect bailout value is always at least as high as
the $k-1$th-step indirect value, and so if it is impossible that $x_i/c_i<1$ for all $i$
under the first-step indirect value
definition of $x_i$, then it is also impossible if we weakly increase the $x_i$'s.\eproof

\bigskip

\noindent{\bf Proof of Proposition \ref{disjointcycles}:} Given that banks lie on at most one cycle,
cycles can be partitioned into tiers.
First, there are cycles that have no directed path coming in from any other cycle --  call
these $\mathcal{C}_0$.  Next, there are cycles that only have directed paths coming in
from $\mathcal{C}_0$, call these $\mathcal{C}_1$.
Then there are cycles that only have directed paths coming in from $\mathcal{C}_0$ and $\mathcal{C}_1$,
and so forth.   Remaining banks either lie on directed paths between cycles,
or on dead-end paths that come out of some cycle(s).   Given that at least one bank must be made
solvent on each directed cycle, it follows that finding the cheapest bank on each cycle in
$\mathcal{C}_0$ is necessary for solvency, and cannot be
made cheaper by any further bailouts.  Under the ordering, these cycles will all be listed first and none
will affect the bailing out of any other within this tier.
Iterating on this logic, the result is easily verified, noting that any bank $i$ that is not on
any cycle will be cleared once all banks that have directed paths leading to it are bailed out,
which necessarily has to be done before any cycles that lie on directed paths that point out from
bank $i$ are bailed out.\eproof
\bigskip

\noindent{\bf Proof of Proposition \ref{star}:} First note that since the network is weakly balanced, and the center bank lies on all cycles, all banks are solvent in the worst equilibrium if and only if the center bank is. Hence the optimal bailout policy is the one ensuring the solvency of the center bank at minimum cost.

The center bank needs an amount of liquidity $(n-1)D^{out}-p_n\leq (n-1)D^{in}$ to be brought back to solvency. Bailing out a peripheral bank $i$ costs $D^{in}-p_i$ and leads to a liquidity flow of $D^{in}$ into the center bank. Hence it is always weakly cheaper (strictly if $p_i>0$) to do so instead of injecting $D^{in}$ directly into the center bank.

The only time at which it may not be optimal to bail out a peripheral bank is then when the center bank is less than $D^{in}$-close to solvency. This is the case if and only if $\lfloor m^*\rfloor$ peripheral banks have already been bailed out. Then the regulator can either bail out one additional peripheral bank, which costs $D^{in}-p_{\lfloor m^*\rfloor+1}$, or inject the center bank's marginal shortfall directly into it. \eproof

\bigskip

\noindent{\bf Proof of Proposition \ref{coreperiphery}:} Let $x\in\{0,1,\dots n_C\}$ be the number of solvent core banks. The following reasoning holds for any $x$. Without loss of generality, suppose there are still some insolvent core banks.

Core banks are completely identical except for the value of their outside assets. Hence bailing out any core bank induces the same liquidity flow into the network, but the cost of such bailout can differ across core banks. It is then always optimal to bring core banks back to solvency in decreasing order of their $p_i$.

What is the cheapest way to bring a core bank $i$ back to solvency? By the previous argument, it cannot be cost-efficient to bailout its core counterparties, as these are more expensive to bailout than $i$. So the optimal way must only involve bank $i$'s peripheral banks, or injecting capital into $i$ directly. This is the same problem as studied in Section \ref{sec:star}, and Proposition \ref{star} applies. The only slight difference is that, for some parameter values, bailing out all of $i$'s peripheral banks may not be enough to ensure its solvency -- i.e. $m_i^*(x)>n_P$. The same logic however still holds: it is still optimal to leverage the peripheral banks' outside assets, and the optimal policy is then to bailout all of $i$'s peripheral banks, and then inject whatever additional capital is needed directly into $i$.    \eproof

\bigskip

\newpage
{\bf Online Supplemental Appendix}

\section{Additional Discussions} \label{appendixB}

\subsection{An Example of a Network with Three Different Equilibria} \label{sec:multiple_ex}

Consider the network depicted in Figure \ref{fig:multple_eq}.

\begin{figure}[!h]
\begin{center}
\begin{tikzpicture}[scale=0.8]
\definecolor{chestnut}{rgb}{0.8, 0.36, 0.36};
\definecolor{afblue}{rgb}{0.36, 0.54, 0.66};
\foreach \Point/\PointLabel in {(6,0)/3, (9,0)/4}
\draw[ fill=afblue!40] \Point circle (0.35) node {$\PointLabel$};
\foreach \Point/\PointLabel in {(3,0)/2, (0,0)/1}
\draw[ fill=chestnut!40] \Point circle (0.35) node {$\PointLabel$};
\draw[->, thick] (0.4,0.4) to [out=40,in=135] node[midway,fill=white]  {$1$} (2.6,0.4);
\draw[<-, thick] (0.4,-0.4) to [out=-40,in=-135] node [ midway,fill=white]  {$1$} (2.6,-0.4);
\draw[->, thick] (3.4,0.4) to [out=40,in=135] node [ midway,fill=white]  {$\frac{3}{4}$} (5.6,0.4);
\draw[<-, thick] (3.4,-0.4) to [out=-40,in=-135] node [ midway,fill=white]  {$\frac{1}{4}$} (5.6,-0.4);
\draw[->, thick] (6.4,0.4) to [out=40,in=135] node[midway,fill=white]  {$1$} (8.6,0.4);
\draw[<-, thick] (6.4,-0.4) to [out=-40,in=-135] node [ midway,fill=white]  {$1$} (8.6,-0.4);
  \end{tikzpicture}
  \end{center}
  \captionsetup{singlelinecheck=off}
  \caption[]{\small Let $p_1=p_4=0$ and $p_2=p_3=1$. }\label{fig:multple_eq}
\end{figure}

Suppose full failure costs, such that a bank's assets are entirely lost upon defaulting: $\beta_i(\textbf{V}, \textbf{p})=p_i + d_i^A(\textbf{V})$.

The best equilibrium has all banks being solvent. Indeed, $p_i+D_i^A-D_i^L\geq 0$ for all bank $i$. Best equilibrium values for banks are then $\mathbf{\overline{V}} = (0, \frac{1}{2},\frac{3}{2},0 )$. The worst equilibrium has all banks defaulting, as none of them is unilaterally solvent: $p_i<D_i^L$ for all bank $i$. Worst equilibrium values for banks are then $\mathbf{\underline{V}} = (-1, -\frac{3}{4},-\frac{1}{4}, -1)$. There exists a third, intermediary, equilibrium in which only banks 1 and 2 default. Indeed, the debt repayment from Bank 3 is not enough to guarantee Bank 2's solvency if Bank 1 defaults on its debt. Bank values in that intermediate equilibrium are $\mathbf{V} = (-1, -\frac{3}{4},\frac{3}{4},0 )$.

\subsection {Complexity Theory: Background Definitions }\label{sec:def}

We present some standard definitions,
included here for the reader who may not be fully familiar with them.\footnote{To keep things
to a reasonable length, we do not include all background definitions of how one defines an algorithm,
how to count its steps, etc., but such background is easy to find in any text on computational
complexity (e.g., \cite{arora2009computational,papadimitriou1994computational}).}

A decision problem -- a problem with a yes/no answer --
that takes as input a network with $n$ banks is in the \textbf{NP}
complexity class if there exists an algorithm 
that can \emph{verify} whether any given
policy is a solution to the problem in
polynomial time, that is there exists a positive constant $r$ such that the
verification algorithm runs in $O(n^r)$-time.
Intuitively, problems in NP are problems for which a solution can be easily verified.

Consider for instance the following decision problem:
\emph{Does there exist a bailout policy that ensures systemic solvency and costs
no more than some budgeted amount $W$?}
This problem belongs to NP: it is easy to check whether a given bailout
policy $(t_i)_i$ ensures systemic solvency and costs no more than $W$,
as this only requires computing $\mathbf{V}(\mathbf{p}+\mathbf{t})$ and $\sum_i t_i$.

A problem is \textbf{NP-hard} if it is at least as hard as any problem in
NP.\footnote{A problem is at least as hard as another if any instance
of the second can be translated into an instance of the first in at most polynomially many steps.
Alternatively, given an oracle that can
solve any instance of that first problem in one step, there exists an algorithm that
solves the latter problem in polynomial time. }
If a problem is NP-hard, then there is no known polynomial-time algorithm that solves it,
and it is believed by many that no such algorithm exists.

A problem is \textbf{strongly NP-hard} if it remains NP-hard even
when the size
of all its inputs is appropriately bounded.\footnote{Many well-studied problems
(e.g., the Knapsack Problem, or the Partition Problem) are only \emph{weakly}
NP-hard, which means that these problems become hard to solve and approximate
only when the size of their inputs becomes large, in a well-defined sense.
For instance the Partition Problem asks whether a given set of $n$ positive
integers can be partitioned into two subsets that have the same sum.  If the
possible integer values that are in the given set are sufficiently limited,
then the problem can become quite easy.  For example, if all these integers
are 0s or 1s, then the problem is trivial as it boils down to checking whether
their sum is even.  More generally, if these integers are bounded above by
some value $cn$, then each subset of the partition can sum to at most $0.5cn^2$,
and an algorithm could check all these possible values exhaustively fairly quickly.
So at least some of the $n$ integer values that are in consideration have to
be ``large'' for the Partition Problem to be hard.
In contrast, the Vertex-Covering Problem that we use in our proof has no integer values associated with it:
it considers an  undirected network on $n$ nodes and asks what is the size of the smallest
set of nodes such that every  link  the network has at least one of its
endpoints in the set.
Importantly, even though there are no integer values associated with the specification of
this problem, the problem can be much harder to solve in that number of potential sets of
nodes that one has to check explodes at a rate of $n!$, and unless the network is special
(e.g., a lattice, bipartite) there are no known simple shortcuts.
}

\subsection{Bailouts versus Guaranteed Payments}

There are two conceptually different ways in which capital
can be injected into a network to return it to solvency:
the regulator can (i) provide sufficient capital to some banks to ensure that they
can pay their debts in full (i.e., bail out some nodes in the network),
or (ii) make some set of payments on behalf of defaulting banks (i.e., pay or guarantee some edges).

Depending on the context, it might be easier to characterize the minimal capital
injections ensuring systemic solvency either in terms of bailouts or guaranteed payments.
Thus, before proceeding, we provide a lemma that outlines the relationship between these two policies.

A \textbf{bailout policy} are bank-specific transfers, $(t_i)_i\in\mathbb{R}_+^n$,
such that the regulator gives capital $t_i$ to bank $i$.
The capital can be given to the banks in any order:
the network will not completely clear until all payments are
made.\footnote{Given the fact that the solvency is independent of the order
of payments, we do not explicitly model the timing of bailouts.
Effectively, a bailout policy will have to make some first set of banks solvent, and then they can repay all of their debts.  Together with the other bailout payments, other banks then become solvent, and make
their payments, etc.}

A set of \textbf{guaranteed payments} is a set of debts ($D_{ij}$s) and associated weights ($\alpha_{ij}$s in $ [0,1]$) such that, for each debt $D_{ij}$, the regulator pays a fraction $\alpha_{ij}$ of it.
These can be tracked as a list of edges in the financial network,  $E\subset n\times n$,
listing the debts to be paid and an associated weight $\alpha_{ij}$ for each $ij\in E$.

\begin{lemma} \label{lem:paymentsvsbailouts}
\
The following hold in any weakly balanced network:
\begin{itemize}
\item[(i)] For any bailout policy that ensures systemic solvency, there exists a set of guaranteed
payments that also ensures systemic solvency and leads to the same total cost.
\item[(ii)] For any set of guaranteed payments that ensures systemic solvency, there exists a bailout policy that also ensures systemic solvency and leads to the same total cost.
\item[(iii)] If the network is exactly balanced, then in order to find
the cheapest policy that ensures full solvency it is without loss of generality to only consider
{full} guarantees on payments (i.e., $\alpha_{ij}\in\{0,1\}$ for all $ij$).
\end{itemize}
\end{lemma}

We do not offer a full proof as it is straightforward,
but simply illustrate Lemma \ref{lem:paymentsvsbailouts} 
via the networks depicted in Figure \ref{fig:balanced}.

\begin{figure}[!h]
\centering
\includegraphics[width=0.8\textwidth]{balanced_examples.tikz}
\caption{\small The network on the left has two cycles: $\{1,2,3, 1\}$ and $\{2,3,4,2\}$.
The network on the right has three cycles: $c_1=\{1,2,1\}$, $c_2=\{3,1,3\}$, and $c_3=\{3,2,1,3\}$.
Arrows point in the direction that debt is owed. }
\label{fig:balanced}
\end{figure}

First, suppose the networks are exactly balanced, and each bank has $p_i=0$.
Recall that, absent any intervention by the regulator,
all banks default in the worst equilibrium since none are unilaterally solvent.
The network on the left has two simple cycles. Making payments $\{D_{12}, D_{34}\}$ ensures that all banks are solvent, and reaches the minimal possible cost of 2. Another way to ensure full solvency at minimum cost is for the regulator to make payment $\{D_{23}\}$.  Note that the first intervention is equivalent to bailing out Bank 4, and then Bank 2.
The second intervention is equivalent to bailing out Bank 3.
Other minimum bailouts in this network are, for instance, just Bank 2,
or Banks 1 and then 4, which respectively correspond to guaranteeing payments
$\{D_{21},D_{24}\}$ and $\{D_{13}, D_{43}\}$.
The network on the right has three simple cycles.
There are three minimum-cost set of payments that ensure systemic solvency:
$\{D_{32},D_{12},D_{31}\}$, $\{D_{12},D_{13}\}$, and  $\{D_{21},D_{31}\}$.
They all lead to a total injection of capital equal to 1.5. They are equivalent to,
respectively: bailing out Bank 2 and then Bank 1;  bailing out Bank 3 and then Bank 2;
bailing out Bank 1. Note that here we only had to consider full guarantees on payments
as the networks are exactly balanced (Lemma \ref{lem:paymentsvsbailouts} (iii)).

Next, suppose each bank has $p_i=.5$.
In both networks, all banks are still insolvent in the worst equilibrium if the regulator does not intervene.
In the network on the left, some of the bailout policies that were minimum cost when the $p_i$s were 0 no longer are.   For instance, bailing out Bank 2 or Bank 3 still ensures systemic solvency, but costs 1.5.
In contrast, bailing out both Bank 1 and  Bank 4 costs 1 in total.  This could be accomplished by making payments $D_{13}$ and $D_{43}$, but the minimum cost would be achieved by paying only a fraction $\alpha_{13}=\alpha_{43}=1/2$ of those debts.  Paying the full debts would double the cost. Hence it is no longer without loss to only consider full guarantees on debt payments, as banks now have some capital buffer.
In the network on the right, there are now two minimum cost bailout sequences instead of three, which consist in  either bailing out Bank 2 or Bank 3. Both require a total injection of 0.5. They are equivalent to, respectively, guaranteeing payment $\{D_{23}\}$ in full and guaranteeing payment $\{D_{31}\}$ with weight $\alpha_{31}=0.5$.  

Despite this equivalence between bailouts and guaranteed payments,
there are still important reasons why a regulator may prefer to
make some of a bank's payments instead of bailing it out.
One is that a bank could use injected capital for purposes other than repaying its debts.
For instance, bailout money was used to pay traders' and bankers' bonuses in the 2008 financial crisis (\cite{story2009}).
Given that such issues are beyond the scope of this paper,
and that bailouts and guaranteed debt payments can be translated into each other in our model,
we move back and forth between the two forms of policy depending on which is
conceptually easier to work with for a given problem.

\subsection{Sufficient Conditions: The Case of Critically Balanced Networks} \label{sec:criticallybalanced}

Although finding the optimal bailout policy is generally complex, there are cases in which a more
precise characterization is possible. Here we consider networks in
which banks have small capital buffers, or none. This makes systemic solvency demanding,
but also provides a bound on minimum bailouts when banks do have capital buffers.

A sufficient condition to ensure the existence of an iteratively strongly solvent
set intersection each cycle is to bailout one bank per cycle.
In some cases, this is also necessary.

Consider any critically balanced network. Let $c_1$, $c_2$, \dots $c_K$ be the list of simple cycles
in the network.  For ease of exposition here,
we deviate from our previous notation and
define a cycle not as a sequence of banks that each owes the next one a debt, but as the list of the corresponding directed edges. Naturally, the two are equivalent.

\begin{proposition}\label{opt_balanced}
Suppose the network is \emph{critically} balanced, and the regulator can only make full payments of debts: $\alpha_{ij}\in\{0,1\}$ for all $i$, $j$.
The minimum capital injections needed in terms of guaranteed payments to ensure full solvency is the
least costly set of payments that includes one on each simple cycle:
\begin{align*}
\min_{E\subseteq N\times N}\quad &\sum_{ji\in E}D_{ij}\\
\text{s.t.}\quad&  E\cap c_k\neq \emptyset \quad \forall k.
\end{align*}
If the network is \emph{exactly} balanced, this is also the optimal policy
when partial payments $\alpha_{ij}\in[0,1]$ are allowed.
\end{proposition}

Recall that in critically balanced networks, one missing debt payment is enough to bring
any bank to insolvency, as a bank's capital buffer is lower than any of its debt assets.
Because all payments are critical, a repayment stream cannot spread from one cycle to another, and the
regulator has to inject some capital into \emph{all} cycles. In exactly  balanced networks, she has to
ensure one payment in full per simple cycle. In critically balanced networks, as some banks may have a
bit of capital buffer, she may only need to make a partial payment.

As soon as some banks have non-trivial capital buffers, such that they
can absorb losses associated with the default of some of their counterparties,
the regulator may not have to inject capital into \emph{all} cycles.
In any case, Proposition \ref{opt_balanced} gives a lower bound on the cost of minimum bailouts,
as additional capital buffer can only help the regulator.

\begin{corollary}\label{opt_bound}
In any network, the minimum total injection of capital needed to ensure full solvency is at most
\begin{align*}
\min_{E\subseteq N\times N}\quad &\sum_{ji\in E}D_{ij}\\
\text{s.t.}\quad&  E\cap c_k\neq \emptyset \quad \forall k
\end{align*}
\end{corollary}

If some banks are not critically balanced, a repayment cascade in one cycle can spread to another,
and bailing out a bank on one cycle can lead other cycles to clear as well.
Figure \ref{fig:bailouts} provides an illustration.

\begin{figure}[!h]
\centering
\includegraphics[width=0.8\textwidth]{balanced_vs_imbalanced.tikz}
\caption{\small $(p_1,p_2,p_3)=(0,0,0)$ in the network on the left,
which is thus exactly balanced. $(p_1,p_2,p_3)=(0,0,D)$ in the network on the right,
which is thus not exactly balanced, nor even critically balanced:
Bank 2 can be made solvent by receiving just the payment from Bank 3.
}\label{fig:bailouts}
\end{figure}

In the exactly balanced network on the left, ensuring solvency of the full network requires bailing
out at least one bank per cycle, which can be done by bailing out Banks 1 and 3, or just Bank 2.
Both policies cost $2D$. Alternatively, the regulator could bailout Bank 1, and then bailout Bank 2
once it has received its debt payment from 1, which would also cost $2D$.
In any case, the regulator has to inject some capital into each cycle to ensure full solvency,
as stated in Proposition \ref{opt_balanced}.

In the network on the right, bailing out Bank 3 is enough to ensure full solvency,
as Bank 3 paying back its debt is enough to make Bank 2 solvent,
which then makes Bank 1 solvent.\footnote{Paying $D/2$ to 1 would not be
enough to get 2 to be solvent since it still gets no payments from 3.
One would need to pay $3D/2$ to 2 in order to ensure its solvency.
So, 3 is the cheapest option.} Hence the regulator does not need to inject capital into each
simple cycle: the repayment cascade in the $\{2,3\}$ cycle spreads to the $\{1,2\}$ cycle without
additional intervention. Importantly this relies on the fact that Bank 2 has a capital buffer:
$p_2+D_2^A=2.5D>1.5D=D_2^L$. 

\subsection{An Upper Bound on Computation Steps }\label{sec:time}

Proposition \ref{NPhard} shows that, as the size of the network $n$ grows,
finding the optimal bailout policy becomes complex.
The intuition is that a network with more banks can have more overlapping cycles,
and these cycles generate interferences
between bailouts. So what drives the complexity of finding the optimal bailout
policy is not
the number of banks $n$ per se, but the associated possible number of cycles $K$.  We formalize
this in Proposition \ref{computation_time}.

\begin{proposition}\label{computation_time}
 The number of calculations needed to find an optimal bailout policy
 is no more than $K!n^{K+1}$, where $n$ is the number of banks in the network
 and $K$ is the number of dependency cycles.
\end{proposition}

\begin{proof}
We first show that there is an optimal bailout policy
that bails out at most one bank per cycle. Although
bailing out one bank per cycle is enough to guarantee systemic solvency (Proposition \ref{solvency}),
this does not imply that bailing out more than one bank per cycle cannot be cost-improving
as it could reduce the cost of a subsequent bailout.

Let $S_0$ be the set of solvent banks in the worst equilibrium absent intervention. Take any optimal bailout policy $(t_i)_i$ that ensures systemic solvency. As explained in Section \ref{minbail}, even though a bailout policy is modeled as a vector of simultaneous payments, it is equivalent to think of it as an ordered list of banks to bailout. Making transfers $(t_i)_i$ leads to a first wave of solvencies of banks $j$ such that $\sum_{k\in S}D_{jk}+t_j\geq D_j^L$. If not all banks are yet solvent, then these new solvencies combined with the transfers lead to a second wave of solvencies, and so forth. Hence a bailout policy defines an order in which banks are made solvent. Relabel banks in that order, such that $i_1$ is the first to be made solvent, then $i_2$, etc. (and if several banks are made solvent in the same wave, break the tie arbitrarily). Since an optimal policy cannot waste transfers, it must set $t_{i_\ell}=\left[D_{i_\ell}^L-\sum_{k\in S\cup\{i_1\}\cup\dots\cup\{i_{\ell-1}\}}D_{i_\ell k}-p_{i_\ell}\right]^+$.

Take any such ordering of banks $(i_\ell)_\ell$ associated with an optimal bailout policy, and
 restrict attention to banks that are actually bailed out, i.e., banks with $t_{i_\ell}>0$.\footnote{If $t_{i_\ell}=0$ then bank $i_\ell$ is not bailed out directly by the regulator but only made solvent by payments it received, at one point or another, from its counterparties.}  Let $\mathcal{K}_\ell$ be the set of cycles on which $i_\ell$ lies. A bailout policy bails out more than one bank per cycle if there exists $\ell$ such
that $\mathcal{K}_\ell\setminus(\cup_{j=1}^{\ell-1}\mathcal{K}_j)=\emptyset$.
That is, at some step, the policy bails out a bank $i_\ell$ that only lies on cycles on which
another bank has already been bailed out.  We show that for any such policy, there exists
another that also ensures systemic solvency, but bails out at most one bank per cycle and
is weakly cheaper. 

Since $i_\ell$ requires a transfer to be solvent despite weak balance, it must be that one of $i_\ell$'s debtors is
defaulting. Let $X_1$ be the set of all  $i_\ell$'s debtors that are defaulting despite already
made bailouts $(i_1,\dots i_{\ell-1})$. It has to be that each bank in $X_1$ also has a debtor
that is defaulting. Iteratively construct $X_k$ has the set of banks that owe money to banks
in $X_{k-1}$ and that are defaulting, and let $X=\cup_k X_k$. Three things are worth noting:
(i) by construction, all banks in $X$ are insolvent, (ii) the subnetwork $(X,(D_{ij})_{i,j\in X})$
must have at least one cycle, and (iii) bank $i_\ell$ is not in $X$. Indeed if $i_\ell\in X$,
it must belong to a cycle of defaulting banks, and hence must belong to a cycle that has not
been cleared out yet, which contradicts our initial assumption. Since the bailout policy
ensures systemic solvency, it has to bailout some banks in $X$ after $i_\ell$'s bailout.
Construct an alternative bailout policy by moving all bailouts of banks in $X$ before $i_\ell$'s bailout.
That cannot increase the cost of these bailouts as banks in $X$ do not expect additional capital
from $i_\ell$ since $i_\ell\notin X$. That makes $i_\ell$'s bailout free -- i.e. unnecessary,-- and
hence strictly reduces the cost of the overall bailout policy. It is thus without loss to only
consider bailout policies that bailout at most one bank per cycle.

We now argue that computing the cost of all such policies requires no more than $K!n^{K+1}$ steps. A bailout policy that bails out one bank per cycle is characterized by (i) the order in which the cycles are cleared, and (ii) the identity of the bank on each cycle that is bailed out. There are $K!$ different orders in which cycles can be cleared. If $m_1$, \dots $m_K$ denote the number of banks on each cycle, there are $K!\prod m_k$ bailout policies to consider. The worst case is when cycles are as large as possible. Since $m_k\leq n$ for all $k$, this means there are at most $K! n^K$ different bailout policies to check. The least costly policy among those must be an optimal policy. Since computing the cost of each policy takes at most $n$ steps, this brute-force algorithm finds the optimal bailout policy in no more than $K!n^{K+1}$ steps.%
\end{proof}

\smallskip
Proposition \ref{computation_time} builds from the following observation:
bailing out more than one bank per cycle cannot be strictly cheaper than
bailing out only one per cycle, and both ensure systemic solvency.
Hence the regulator can restrict attention to
policies that bail out at most one bank per cycle.  Given potential overlap in cycles and
resulting cascades, one needs to do more than just pick the cheapest bank per cycle.
A brute-force algorithm to find the optimal bailout
policy is then to compute the cost of all such policies, and pick the cheapest one.
There are  $K!$ different orders in which the cycles can be cleared, with at most $n$ different
banks per cycle, leading to at most $K!n^K$ policies (a bank per cycle and the order of their clearance).
The computation of the cost of each policy
takes at most $n$ steps, and so this brute force algorithm can take at most this many
steps (see the appendix for a formal proof). In networks with large cycles,
an exhaustive search approaches this number of steps.

Proposition \ref{computation_time} implies that when the number of cycles is small, the problem remains tractable even if the number of
banks is very large.  For instance, if there is some bounded number of cycles so that $K$ is
small and fixed, then the brute-force
algorithm described above runs in $O(n^{K+1})$-time as $n$ becomes large.
In contrast, finding the optimal
bailout policy is most complex when there are many overlapping cycles as for instance in cliques,
and we discuss those in Section \ref{sec:core-periphery}.  In settings with cliques, $K$ becomes
exponential in the size of the cliques and thus can become exponential in $n$.

\subsection{A $2\times 2$ Core-Periphery Network with Asymmetries}\label{asyexample}

Consider a network with two core banks $i=2,3$, and two peripheral banks $i=1,4$. Core banks have claims on each other, and a claim on their peripheral bank. Each peripheral bank has a claim on its core bank. The financial network thus contains three simple cycles of claims---one between the two core banks, and one between each core bank and its peripheral bank---as depicted in Figure \ref{fig:asymmetriccoreperi}.
\begin{figure}[!h]
\begin{center}
\begin{tikzpicture}[scale=0.8]
\definecolor{afblue}{rgb}{0.36, 0.54, 0.66};
\definecolor{chestnut}{rgb}{0.8, 0.36, 0.36};
\foreach \Point/\PointLabel/\Color in {(0,2)/1/afblue, (3,2)/2/chestnut, (6,2)/3/chestnut, (9,2)/4/afblue}
\draw[fill=\Color!40] \Point circle (0.35) node {$\PointLabel$};
\draw[->, thick] (0.4,2.4) to [out=40,in=135] node[midway,fill=white]{$D_{21}$} (2.6,2.4);
\draw[<-, thick] (0.4,1.6) to [out=-40,in=-135] node[midway,fill=white]{$D_{12}$}  (2.6,1.6);
\draw[<-, thick] (3.4,2.4) to [out=40,in=135] node[midway,fill=white]{$D_{23}$}  (5.6,2.4);
\draw[->, thick] (3.4,1.6) to [out=-40,in=-135] node[midway,fill=white]{$D_{32}$} (5.6,1.6);
\draw[<-, thick] (6.4,2.4) to [out=40,in=135] node[midway,fill=white]{$D_{34}$}  (8.6,2.4);
\draw[->, thick] (6.4,1.6) to [out=-40,in=-135] node[midway,fill=white]{$D_{43}$} (8.6,1.6);
  \end{tikzpicture}
  \end{center}
  \captionsetup{singlelinecheck=off}
  \caption[]{Arrows point in the direction that debt is owed, that is from debtors to creditors.
  There are two core banks (pink) and two peripheral banks (blue).}
  \label{fig:asymmetriccoreperi}
\end{figure}
The network is weakly-balanced, so all banks are solvent if they get their debt paid back in full.

Suppose that peripheral banks are ``small,'' in the sense that them repaying back their
debt is never enough to ensure the solvency of their core bank.\footnote{Formally,
$p_2+D_{21}<D_2^L$ and $p_3+D_{34}<D_3^L$.} Then it is always cheaper to start by bailing
out a peripheral bank before bailing out its core bank.
Perhaps counterintuitively, it is only when a peripheral bank can induce a
cascade that it might not be optimal to bail it out. Indeed, if the peripheral
bank's payment to the core bank is more than enough to make the latter solvent,
then the regulator might ``overpay'' by bailing out the peripheral bank because the cost of bailout
might be more than the shortfall in the core bank.
However, if a peripheral bank is small, then the regulator cannot be
overpaying by bailing it out first, but can only gain from doing so.

From now on, assume that peripheral banks are small. Then the optimal bailout policy must start
by bailing out a peripheral bank. There are then four candidate policies:
$(1,2,3)$, $(1,2,4)$, $(4,3,2)$, and $(4,3,1)$. Even in this simple network, there is
no systematic method to determine in which order core banks should be made solvent
(the first two policies make Bank 2 solvent first, and then Bank 3, whereas the last
two policies do the reverse.) What is true, however, is that making a core bank a larger
debtor (i.e., increasing its $D_i^L$) while keeping its shortfall $D_i^L-p_i$ constant always
favors bringing that core bank back to solvency before other core banks.
That is, if under some
initial $\mathbf{D}, \mathbf{p}$, the optimal bailout policy makes Bank 2 solvent before Bank 3,
then that order remains optimal if $D_2^L$ and $p_2$ are both increased by the same amount.
Indeed, this increases the indirect bailout value of Bank 2 without changing how much liquidity it
needs, making it more likely that bailing it out before Bank 3 is optimal.

\subsection{Incorporating Equity-Like Interdependencies Between Financial Institutions} \label{sec:equity}

In this section, we propose an extension of the model that includes both debt and equity.

The distinction between debt and equity is not just a theoretical consideration, since both types of securities are needed to capture the balance sheets of some of the most prominent and important types of financial institutions.
For example, banks' balance sheets involve substantial portions of deposits,
loans, CDOs (collateralized debt obligations), and other sorts of
debt-like instruments.    In contrast, venture capital firms and many other
sorts of investment funds typically hold equity, and are either held
privately or issue equity.
Furthermore, some large investment banks are hybrids that involve substantial portions of both types of
exposures.
Finally, our model also captures interactions within financial conglomerates, as organizations
within the same group often have both debt-like and equity-like exposures
to each other (e.g., see the report of the Basel Committee on Banking Supervision \citeyearpar{BISreport_conglomerates}
for a description of intra-group exposures).
Understanding the different incentives these forms of
interdependencies provide, and the externalities they generate,
is thus relevant.\footnote{There are obviously more complex contracts that can also be built into such a model.
In Section \ref{generalf} of the online materials we consider a model allowing for general contracts, and discuss how existence of consistent bank values depends on the monotonicity of those contracts in underlying primitive-asset investments. Debt and Equity  provide a lens into many other contracts, as many swaps and derivatives
involve either fixed payments or payments that depend on the realization of the
value of some investment of one of the parties, like a combination of debt and equity.}

\paragraph{Equity Contracts.} Interdependencies can now also take the form of equity contracts. If bank $i$ owns an equity share in bank $j$, it is represented by a claim that $i$ has on some fraction $S_{ij}\in [0,1]$ of $j$'s value. A bank cannot have an equity claim on itself, so that $S_{ii}=0$ for all $i$. Equity shares must sum to one, so whatever share is not owned by other banks accrues to some outside investor:
$S_{0i}=1-\sum_{j\neq 0} S_{ji}$.\footnote{Here, we simply model
any fully
privately held banks as having some outside investor owning
an equity share equal to 1.
This has no consequence, but allows us to trace where
all values ultimately accrue.}
The one exception is that no shares are held in the outside investors so that $S_{i0}=0$ for all $i$ -- shares held by banks in private enterprises are modeled via the $p_i$'s.

Finally, in order to ensure that the economy is well-defined,
we presume that there exists a
directed equity path from every bank to some private investor (hence to node 0).
This rules out nonsensical cycles where each bank is {\sl entirely}
owned by others
in the cycle, but none are owned in any part by any private investor.
For instance
if A owns all of B and vice versa, then there is no solution to equity values.

\paragraph{Bank Values.}  The value  $V_i$ of a bank $i$ equals:
\begin{align} \label{eqn:Vequity}
V_i=p_i +\sum_{j} S_{ij} V_j^+  +   d_j^A(\mathbf{V})-D^L_i  - b_i(\mathbf{V},\mathbf{p}),
\end{align}
where $V_j^+\equiv\max [V_j, 0]$. The $S_{ij} V_j^+$ reflects limited liability: the value to $i$ of its equity holding in
$j$ cannot be negative. failure costs $b_i(\mathbf{V},\mathbf{p})$ are defined by:
\begin{equation*}
  b_i\left(\mathbf{V},\mathbf{p}\right) =
  \begin{cases}
   0 & \text{ if  }p_i+\sum_{j} S_{ij} V_j^+  +d_i^A(\mathbf{V})  \geq D_i^L \\
   \beta_i\left(\mathbf{V},\mathbf{p}\right) & \text{ if  } p_i+\sum_{j} S_{ij} V_j^+  +d_i^A(\mathbf{V})  <D_i^L.
  \end{cases}
\end{equation*}

In matrix notation, equation (\ref{eqn:Vequity}) becomes
\begin{equation}\label{V-value}
\mathbf{V}=(\mathbf{I}-\mathbf{S}(\mathbf{V}))^{-1} \left(\left[ \mathbf{p} + \mathbf{d}^A(\mathbf{V})  - \mathbf{D}^L\right] - \mathbf{b}(\mathbf{V},\mathbf{p})\right),
\end{equation}
where  $\mathbf{S}(\mathbf{V})$ reflects the fact that $S_{ij}(\mathbf{V}) = 0 $ whenever $j$ defaults, and
equals $S_{ij}$ otherwise.

\paragraph{The Existence of Values Satisfying Equation (\ref{V-value})} $(\mathbf{I}-\mathbf{S})$ is invertible if (and only if) the matrix power series
$\sum_{k=0}^\infty \mathbf{S}^k$ converges, which is equivalent to
the largest eigenvalue of $\mathbf{S}$, in magnitude, being strictly below one.
Let us treat the case in which $\mathbf{S}\neq 0$, as otherwise the result is obvious.
Denote by $\lambda$ the largest eigenvalue in magnitude, and $w$ the associated eigenvector.
From the Perron Frobenius theorem, we know that $\lambda\geq 0$ and $w$ is nonnegative and nonzero.

By contradiction, suppose that $\lambda\geq 1$. Then $\sum_{j\neq 0} S_{ij}w_j =
\lambda w_i $ for each $i$\footnote{Recall that $S_{i0}=0$ for all $i$.}
implies that $\sum_{j\neq 0} w_j \sum_{i\neq 0} S_{ij}\geq \sum_{i\neq 0}w_i$.
Since $\sum_{i\neq 0} S_{ij}\leq 1$ for all $j$, this is equivalent to
$\sum_{j\neq 0} w_j \sum_{i\neq 0} S_{ij}= \sum_{i\neq 0}w_i$.
To ease the comparison of the LHS and RHS, rewrite the indices on the left side as
$\sum_{i\neq 0} w_i \sum_{j\neq 0} S_{ji}= \sum_{i\neq 0} w_i$.\footnote{Note that this does not change anything, as we are still summing over the same set.}

This requires that if $w_i>0$, then $\sum_{j\neq 0} S_{ji}=1$.
Since the eigenvector is not all zeros, we know there exists at least one bank $i$ with $w_i>0$.
If $i$ is such that $S_{0i}=1- \sum_{j\neq 0} S_{ji}>0$, then we get a contradiction directly.
If instead $i$'s equity value is entirely owned by other financial institutions, there
must exist another bank $j$ with $S_{ji}>0$. This implies $w_j>0$.
Same argument applies: either $j$ is partly owned by outside investors, in which case
we directly get a contradiction, or we can move back the equity path to, yet another, bank.
Since there must exist an equity path from outside investors to any bank, this process must
terminate to some bank $j'$ that is, at least partly owned by outside investors, such that it
has $\sum_{j\neq 0} S_{jj'}<1$ and yet $w_{j'}>0$.
This is a contradiction of the inequality we
started from, and hence $\lambda<1$.

\paragraph{Consistency of Bank Values}

Let us check that the values of the banks are actually consistent
in adding up appropriately to the total value of the portfolio of investments.
For simplicity we analyze the case without failure costs.
then in terms of matrix notation,  bank values solve
\[\mathbf{V}= \mathbf{q} \mathbf{p}  + \mathbf{D}^A  - \mathbf{D}^L + \mathbf{S}\mathbf{V}^+.\]
Written this way, the book or equity value of a publicly held organization coincides with its total market value.
Indeed as argued by both Brioschi, Buzzacchi, and Colombo \citeyearpar{brioschibc1989} and  Fedenia, Hodder, and Triantis \citeyearpar{fedeniaht1994}, the ultimate (non-inflated) value of an organization to the economy -- what we call the ``market'' value --  is well-captured by the equity value of that organization that is held by its \emph{outside} investors -- or the {\sl final} shareholders who are private entities that have not issued shares in themselves. This value captures the flow of real assets that accrues to final investors of that organization.
This is exactly what is characterized by the above values since summing them up (again, for the case of nonnegative values) gives
\begin{align*}
\sum_{i\neq 0}V_i&=\sum_{i\neq 0}\sum_k q_{ik} p_k+ \sum_{i\neq 0}  D_i^A -\sum_{i\neq 0} D_i^L + \sum_{i\neq 0}\sum_{j\neq 0} S_{ij} V_j \\
& =\sum_{i\neq 0}\sum_k q_{ik} p_k + D_0^L -D_0^A + \sum_{j\neq 0}(1- S_{0j}) V_j \\[7pt]
\implies D^A_0&-D^L_0+ \sum_{i}S_{0i}V_i =\sum_i\sum_k q_{ik} p_k
\end{align*}
It is easy to see that the total equity value accruing to all private investors
(so value net of debt)
equals the total value of primitive investments.\footnote{Note that
when debts are zero, this value ends up being the same as that
in (3) of
Elliott, Golub and Jackson \citeyearpar{elliottgj2014}.  The difference is that here
we explicitly model the outside investors as being part of the network,
which enables us to simplify the solution, eliminating the need
for tracking the $\widehat{C}$ matrix that was used there.}

\subsubsection{Extension of our Results with Equity Claims}

\paragraph{Equilibrium Multiplicity.} In our benchmark model, Proposition \ref{uniqueV} states that any cycle of debt claims can lead to multiple equilibria for bank values. Reciprocally, without any cycle of debt, equilibrium values must be unique. This is no longer true when banks can also be linked via equity contracts. A cycle composed of a mix of equity and debt claims is enough to generate multiple equilibria. A cycle composed solely of equity claims is not, however. Hence, Proposition \ref{uniqueV} extends by defining a dependency cycle as a cycle of claims that involves \emph{at least some} debt.

\paragraph{Minimum Bailouts.} The general program to ensure solvency at minimum cost can be written as
\[
\min_{p'\geq p:   \mathbf{V}(\mathbf{p}')\geq 0}  ||\mathbf{p}'-\mathbf{p}||,
\]
where the $\mathbf{V}$
is chosen to be either the best or worst equilibrium, depending on which is of
interest.
Again, this requires that the imbalance is at most 0 for all banks, but now this includes equity values and so has to be solved as a fixed point.

The algorithm for finding the amount needed to remove the net imbalance in the case of the best equilibrium is straightforward to describe.
It is as follows.

Let
\[
p_i^n= p_i + D_i^A-D_i^L.
\]
Now, calculate the best equilibrium values associated
with asset returns $\mathbf{p}^n$,
equity holdings $\mathbf{S}$
(noting that equity values in negative-valued enterprises are 0),
and no debt $\mathbf{D}=\mathbf{0}$.
The opposite of the total sum of the valuations of the banks with negative values (ignoring failure costs)
is the minimum bailout that is needed.
Effectively, we know that all debts will be repaid in a bailout that ensures
full solvency, and then the resulting
bank values
will be the basis on which equity values accrue.  Banks that are still negative,
including all of their equity positions, are the ones that will require bailout payments.

In the case of the worst equilibrium, the same logic applies, but then the
base values are associated with the worst equilibrium.
Then once those payments are made, one {\sl recalculates} the worst
equilibrium values given those payments, but with the original $\mathbf{D}$.
By doing this, one identifies banks that are then unilaterally solvent
(after the initial bailout payments),
as well as any resulting iteratively strongly solvent set by consequence of those unilateral solvencies.
If these are not enough to intersect each directed cycle, then additional bailouts will be needed, and an algorithm needs to be run to find the cheapest set.  Note that those bailouts might not even be used to generate unilateral solvencies, but might just be enough to generate secondary solvencies given the unilateral solvencies, which eventually generate more solvencies.   This is the analog of the problem without equity, but just augmented by additional value calculations that include equity of the resulting solvent banks for each possible configuration of bailouts that is considered and the corresponding worst equilibria.
If one can
compress the network, then the issues with the worst
equilibrium are avoided and one only
has to deal with the initial bailouts needed to restore weak balance, which are necessary in any case.

\subsection{General Contracts Between Financial Institutions}
\label{generalf}

We now discuss bank values when contracts are not restricted to debt and equity.
A general form of contract between institutions $i$ and $j$ is denoted by $f_{ij}(\mathbf{V},\mathbf{p})$ and can depend on the value of institution $j$ as well as the value of other institutions. This represents some stream of payments that $j$ owes to $i$, usually in exchange for some good, payments, or investment that has been given or promised from $i$ to $j$.


The value $V_i$  of a bank $i$
is then
\begin{equation}
V_i=p_i+ \sum_j  f_{ij}(\mathbf{V},\mathbf{p}) - \left[ \sum_j  f_{ji}(\mathbf{V},\mathbf{p})- S_{ji}(\mathbf{V})V_i^+\right] - b_i(\mathbf{V},\mathbf{p}), \label{eqn:V1}
\end{equation}
where $ f_{ji}(\mathbf{V},\mathbf{p})- S_{ji}(\mathbf{V})V_i^+$ accounts for the fact that
debt and contracts other than equity are included as liabilities in a book value calculation.\footnote{This more general model also embeds that of Barruca et al. \citeyearpar{barucca2016} in which banks hold debt on each other, but these debt claims are not valued under full information: they allow for uncertainty regarding banks' external assets and ability to honor their interbank liabilities, whose face value may then be discounted depending on available information.   Financial contracts as defined here can capture this kind of uncertainty if $f_{ij}$ equals the expected payment from $j$ to $i$ given some information---e.g. a subset of known bank values or primitive asset values.}

Under monotonicity assumptions on financial contracts, there exist consistent values for
banks by Tarski's fixed point theorem.
That is, there exists a fixed point to the above system of equations.
This is true whenever $b_i(\mathbf{V},\mathbf{p})$ is nonincreasing in $\mathbf{V}$ and
bounded (supposing that the costs cannot exceed some total level),  each $f_{ij}(\mathbf{V},\mathbf{p})$
is also nondecreasing in $\mathbf{V}$, $\sum_j  f_{ji}(\mathbf{V},\mathbf{p})- S_{ji}(\mathbf{V})V_i$
is nonincreasing in $\mathbf{V}$, and either $f$ is bounded or possible values of $\mathbf{V}$ are bounded.
Moreover, from Tarski's theorem,
it also follows that the set of equilibrium bank values forms a complete lattice. Discontinuities, which come from failure costs and
potentially the financial contracts themselves, can thus lead to multiple solutions for banks' values.

\begin{figure}[!h]
 \begin{center}
    \begin{subfigure}[b]{0.48\textwidth}
 \includegraphics[width=\textwidth]{contracts.tikz}
\caption{Non-Decreasing Financial Contracts}
     \end{subfigure}
     \hspace{0.1cm}
    \begin{subfigure}[b]{0.48\textwidth}
 \includegraphics[width=\textwidth]{contracts2}
\caption{Non-Increasing Financial Contracts}
     \end{subfigure}
\end{center}
\end{figure}

When financial contracts are not increasing functions of $\mathbf{V}$,
there may not exist an equilibrium solution for bank values.
For instance, as soon as some banks insure themselves against
the default of a counterparty or bet on the failure of another, simple accounting rules may not yield consistent values for all organizations in the financial network. We illustrate this in the following example.

\paragraph{Example of Non-Existence of a Solution for $\mathbf{V}$: Credit Default Swaps.}
Consider a financial network composed of $n=3$ banks. For simplicity their portfolios in outside assets all have the same value $p_i=2$ for $i=1,2,3$.
The values of banks
are linked to each other through the following financial contracts: bank 2 holds debt from 1 with face value $D_{21}=1$; 2 is fully insured against 1's default through a CDS with bank $3$ in exchange of payment $r=0.4$; finally 1 holds a contract with 3 that is linearly decreasing in 3's value. Suppose an bank defaults if and only if its book value falls below its interbank liabilities, in which case it incurs a cost $ \beta = 0.1$. Formally, the contracts are
\begin{align*}
f_{21}(\mathbf{V}) = D_{21}\mathbbm{1}_{V_1\geq  0}\\
f_{23}(\mathbf{V}) = D_{21}\mathbbm{1}_{V_1<  0}\\
f_{32}(\mathbf{V}) = r\mathbbm{1}_{V_1\geq  0}\\
f_{13}(\mathbf{V}) = -0.5V_3.
\end{align*}
Note that banks 2 and 3 never default: the former's value is always at least $2-r>0$ and
the latter's is at least $2- D_{21}>0$. We then check that there is no solution in which bank 1
is solvent. In such a case, $V_3 = 2+ r$ and $V_1 = 2 -0.5 V_3 - D_{21} = -0.2<0$: but then bank 1 defaults, which
is a contradiction. Finally suppose that 1 defaults.
Then $V_3 = 2- D_{21}$ and $V_1 = 2-0.5V_3-\beta = 1.4>0$, another contradiction.

\end{appendices}
\end{spacing}
\end{document}